\documentclass[aps,prd,showpacs,twocolumn,amsmath,amssymb]{revtex4-1}
\usepackage{epsfig}
\usepackage{graphicx}
\usepackage{dcolumn}
\usepackage{bm}
\DeclareMathOperator{\acot}{acot}
\begin{document}

\author{Guillermo A. Gonz\'alez}
\email[]{guillermo.gonzalez@saber.uis.edu.co}
\author{Oscar M. Pimentel}
\email[]{oscar.pimentel@correo.uis.edu.co}
\affiliation{Escuela de F\'isica, Universidad Industrial de Santander, A. A. 678, Bucaramanga 680002, Colombia}

\title{Static Thin Disks with Haloes as Sources of Conformastatic Spacetimes}

\begin{abstract}

Two new families of exact solutions to the Einstein equations for a conformastatic spacetime with axial symmetry are presented which describe thin disks of dust immersed in a spheroidal halo. The solutions are obtained by expressing the metric function in terms of an auxiliary function which satisfies the Laplace equation, a characteristic property of the conformastatic spacetimes. The first family of solutions is obtained from the displacement, cut and reflexion method, which introduces a discontinuity in the first $z$-derivate of the metric tensor across the plane of the disk. The second family of solutions is obtained by using the oblate spheroidal coordinates because they adapt to the shape of the source and introduce naturally a cutting radius for the disk. The energy densities of the disk and the halo are everywhere positive and well behaved and their energy-momentum tensor agrees with all the energy conditions. Some particular solutions for the energy density of the disk and the halo are presented and the rotational curves are obtained by solving the geodesic equation for a particle that moves in circular orbits in the plane of the disk.

\end{abstract}


\maketitle

\section{\label{sec:intro}Introduction }

The application of the general relativity on the description of astrophysical objects has been matter of interest in the theoretical physics. Exact solutions to the Einstein field equations with spherical symmetry are of great relevance if systems like stars, planets or spherical clusters are considered. Now, in order to model systems like galaxies, accretion disks or rotating black holes, axially symmetric solutions must be considered. For the case of galaxies, the first disks models in general relativity were proposed by Bonnor and Sackfield in 1968 \cite{15} and the second one by Morgan and Morgan in 1969 \cite{16}. These solutions increased the interest in this kind of solutions, and later, many other solution were proposed: solutions for static sources are presented in \cite{17, 18, 19, 20}, while disks with a rotational motion (stationary) are presented in \cite{21, 22, 23}.

The study of the gravitational field produced by an astrophysical source, from the point of view of general relativity, has been a difficult task due to the nonlinear nature of the Einstein equations. The situation is not better when we try to find solutions that describe the gravitational field of a system consisting of two or more parts. Examples of this kind of systems are the galaxies because they have (in general) a disk, a galactic halo, a bulb, and in many cases, a central black hole. The case of a disk immersed in a galactic halo is interesting because the halo is the largest and most massive part of a galaxy. This problem has already been treated by Vogt and Letelier in \cite{24} and by Guti\'errez, Gonz\'alez and Quevedo in \cite{25}. In the first paper the authors apply the displacement, cut and reflect method to some known disk-like solutions in order to generate disks with haloes, and in the second paper the authors present a new family of disk-like solutions with halo for a conformastatic spacetime in Einstein-Maxwell gravity.

In this paper we present two families of axially symmetric thin disks of dust immersed in a spheroidal halo. The disks are obtained from solutions of the Einstein equations for an axially symmetric conformastatic spacetime in which the metric tensor is characterized only by one metric function. By introducing a finite discontinuity on the first derivatives of the metric tensor, solutions with a singularity of the delta function type are obtained, so describing thin disks. The nonzero components of the energy-momentum tensor, both for the disk and the halo, are obtained from the Einstein equations. In this way, the energy densities and pressures of the sources are determined. By imposing the fulfillment of all the energy conditions we obtain a constraint over the solutions, in such a way that the metric function can be properly expressed in terms of a solution of the Laplace equation. By using the solution to Laplace equation in cylindrical coordinates we find infinite disks and by using the solution to Laplace equation in oblate spheroidal coordinates we find finite disks. In both cases we obtain particular solutions with energy densities and pressures well behaved everywhere. We also show that the masses of de disks and the haloes are finite. Finally we solve the geodesic equation for circular orbits in the plane of the disk. Some preliminaries results of this work have been presented in \cite{14}.

The structure of this article is the follow: In Section \ref{sec:sols} we solve the Einstein equations for a conformastatic metric in order to find the energy density and stress of the halo in terms of solutions to the Laplace equation. In Section \ref{sec:emtensor} we apply the thin shell formalism to find a disk-like source that satisfy the energy conditions (physically well behaved). In Section \ref{sec:mass} we introduce the gravitational mass, both for the disk and for the halo, and we study the dynamics of a particle moving in circular orbits in the plane of the disk in order to obtain the rotational curves, which describe the radial dependency of the speed of a particle that moves in circular paths around the center of the galaxy and in the plane of the disk. We then present, in Section \ref{sec:kuzmin}, a particular family of relativistic solutions of infinite thin disks with halo and we analyze the behavior of the physical properties and, in Section \ref{sec:morgan}, we present a second family of solutions of finite thin disks with halo and discuss the physical features of this solutions.  Finally, in Section \ref{sec:concluding}, we present a brief discussion of our main results.

\section{\label{sec:sols}Axially symmetric solutions to the Einstein equations}

In order to formulate the Einstein field equations for an axially symmetric static spacetime with an infinitesimally thin discoidal source which is immersed in a fluid or matter halo, we first introduce the cylindrical coordinates $x^{\alpha}=(t,r,\varphi,z)$. Now, we consider the spacetime as described by the conformastatic metric \cite{1}
\begin{equation}
ds^{2}=-e^{2\psi}dt^{2}+e^{-2\psi}(dr^{2}+r^{2}d\varphi^{2}+dz^{2}),
\label{eq1}
\end{equation}
where we demand $\psi$ to be a continuos function of the $r$ and $z$ coordinates. The Einstein field equations for this metric yield the energy-momentum tensor for the halo fluid $T_{\alpha\beta}$, whose non-null components are
\begin{subequations}\label{E0116}
\begin{eqnarray}
T_{00}&=&e^{4\psi}(2\nabla^{2}\psi-\nabla\psi\cdot\nabla\psi),\\
T_{11}&=&\psi_{,z}^{2}-\psi_{,r}^{2},\\
T_{22}&=&r^{2}\nabla\psi\cdot\nabla\psi,\\
T_{33}&=&-T_{11},\\
T_{13}&=&T_{31}=-2\psi_{,r}\psi_{,z}.
\end{eqnarray}
\end{subequations}  

Now, to give a physical interpretation to the energy-momentum tensor of the halo, we write $T_{\alpha\beta}$ in a locally minkowskian reference frame whose space-like vectors are oriented in the main stresses directions. This reference frame is described by the orthonormal basis,
\begin{subequations}\label{E0116}
\begin{eqnarray}
\zeta_{\hat{0}}^{\alpha}&=&e^{-\psi}(1,0,0,0),\\
\zeta_{\hat{1}}^{\alpha}&=&\frac{e^{\psi}}{r}(0,0,1,0),\\
\zeta_{\hat{2}}^{\alpha}&=&\frac{e^{\psi}}{\sqrt{1+\Omega^{2}}}(0,1,0,-\Omega),\\
\zeta_{\hat{3}}^{\alpha}&=&\frac{e^{\psi}}{\sqrt{1+\Omega^{2}}}(0,\Omega,0,1),
\end{eqnarray}
\end{subequations}
where $\Omega=\psi_{,r}/\psi_{,z}$. By writing $T_{\alpha\beta}$ in this reference frame, we find the energy density
\begin{equation}
\rho = e^{2\psi}(2\nabla^{2}\psi-\nabla\psi\cdot\nabla\psi),\label{rho}
\end{equation}
the main stresses of the halo, which are given by
\begin{eqnarray}
p_{1} &=& e^{2\psi}\nabla\psi\cdot\nabla\psi,\\
p_{2} &=& e^{2\psi}\nabla\psi\cdot\nabla\psi,\\
p_{3} &=& -e^{2\psi}\nabla\psi\cdot\nabla\psi,
\end{eqnarray}
and the isotropic pressure
\begin{equation}
p = \frac{1}{3}e^{2\psi}\nabla\psi\cdot\nabla\psi,
\end{equation}
which is defined as the average of the main stresses.

The physical quantities we have already obtained depend on the unknown metric function $\psi$, so we need an additional equation to get $\psi$. On the other hand, if we demand $\rho$ and $p$ to satisfy the energy conditions \cite{2}, then we get the inequality
\begin{equation}
\nabla^{2}\psi\ge\nabla\psi\cdot\nabla\psi.
\end{equation}
These conditions are important to yield physically well behaved solutions, in other words, to fulfill the minimum conditions for gravitational systems. One way to satisfy (7) is considering $\psi$ functions that are solutions to the equation
\begin{equation}
\nabla^{2}\psi=k\nabla\psi\cdot\nabla\psi,
\label{eq8}
\end{equation}
with $k\ge1$. This particular choice to fulfill the energy conditions leads us to the state equation
\begin{equation}
p=\frac{\rho}{3(2k-1)}
\label{eq9}
\end{equation}
for the halo. Here we can see that the pressure and the energy density are linearly related by a constant $\gamma=1/3(2k-1)$ that depends on $k$ value. It means that a specific kind of fluid in the halo depends on a particular choice of $k$; for example, if $k$ tends to infinity, then $p$ tends to zero and the fluid in the halo would be made of dust. If we choose $k$ equal to 1, then the state equation would describe a fluid made of radiation. Henceforth, we will not analyze the behavior of $p$, because it is related $\rho$ via the state equation.

Now, in order to get solutions for the $\psi$ function, we rewrite equation (\ref{eq8}) as
\begin{equation}
\nabla^{2}(e^{-k\psi}) = 0,
\end{equation}
in such a way that solutions for $\psi$ can be obtained form solutions of the Laplace equation. Furthermore, as the spacetime needs to be asymptotically flat, then $e^{-k\psi}$ has to be $1$ at infinity, and we can write \cite{3}
\begin{equation}
e^{-k\psi} = 1 - U, \label{eq10}
\end{equation}
where $U$ es any solution of the Laplace equation that vanishes at infinity.

This relation allows us to write $\rho$ in terms of $U$ as
\begin{equation}
\rho=\frac{2k-1}{k^{2}}\frac{U_{,r}^{2}+U_{,z}^{2}}{\left(1-U\right)^{2(k+1)/k}}.
\label{eq10u}
\end{equation}
It is clear that we can obtain different solutions depending on the choice of $U$; nevertheless, in the next section we will see that $U$ must satisfy some conditions in order to have a disk solution. 

\section{\label{sec:emtensor}The Energy-Momentum Tensor of the Disk}

In the preceding section we computed the energy-momentum tensor for the fluid in the halo; now, we want to compute the one of the disk. To accomplish this objective we use the distributional approach \cite{6} and the thin shells formalism \cite{7, 8} to describe the disk-like source. Mathematically the disk is described by an hypersurface $\Sigma$ whose function is $\Phi(x^{\alpha})=z$ and its normal vector is $n_{\alpha}=\delta_{\alpha}^{z}$. 
This hypersurface separates the spacetime in two regions: $M^{+}$ (above $\Sigma$) and $M^{-}$ (bellow $\Sigma$). By using distributional functions, the energy-momentum tensor can be written as
\begin{equation}
T_{\alpha\beta}=T_{\alpha\beta}^{+}\theta(z)+T_{\alpha\beta}^{-}\{1-\theta(z)\}+Q_{\alpha\beta}\delta(z),
\label{eq13}
\end{equation}
where $\theta(z)$ is the Heaviside function, $\delta(z)$ is the Dirac function, $T^{\pm}_{\alpha\beta}$ is the energy-momentum tensor in $M^{\pm}$, and $Q_{\alpha\beta}$ is the energy-momentum tensor in $\Sigma$.

Now, to describe properly a surface layer it is necesary to satisfy the condition of continuity of the metric across the plane $z=0$, which is mathematically expressed via the jump function as
\begin{equation}
\left[g_{\alpha\beta} \right]=g_{\alpha\beta}|_{_{z=0^{+}}}-g_{\alpha\beta}|_{_{z=0^{-}}}=0.
\label{eq14}
\end{equation}
This condition can be set in terms of the solution to Laplace equation through (\ref{eq1}) and (\ref{eq10}) as
\begin{equation}
U(r,z)=U(r,-z).
\label{eq15}
\end{equation}
We also need to ensure that $g_{\alpha\beta,z}$ is discontinuous in the plane $z=0$, i.e.
\begin{equation}
b_{\alpha\beta}=\left[g_{\alpha\beta,z} \right]=2g_{\alpha\beta,z}|_{_{z=0^{+}}}\neq 0.
\label{eq16b}
\end{equation}
With the metric tensor defined in (\ref{eq1}) and the relation (\ref{eq10}) it is easy to show that the last condition is satisfied if we choose a function $U$ such that
\begin{equation}
U_{,z}|_{_{z=0^{+}}}\neq 0
\label{eq17u}
\end{equation}

On the other hand, we can compute $Q_{\alpha\beta}$ through the expression
\begin{equation}
Q_{\alpha\beta}=H_{\alpha\beta}-\frac{1}{2}g_{\alpha\beta}H,
\label{eq16}
\end{equation}
where
\begin{equation}
H_{\alpha\beta}=\frac{1}{2}\{b_{\beta}^{z}n_{\alpha}-b_{\nu}^{\nu}n_{\alpha}n_{\beta}+b_{\alpha z}n_{\beta}-b_{\alpha\beta}g^{zz}\}
\label{eq17}
\end{equation}
is the part of the Ricci tensor associated to $\Sigma$. Here it is important to note that $Q_{\alpha\beta}\neq0$ by virtue of the condition (\ref{eq16b}). Computing $Q_{\alpha\beta}$ we find that its only non-null component is
\begin{equation}
Q_{t}^{t}=-4\psi_{,z}e^{2\psi},
\label{eq18}
\end{equation}
which is evaluated in $z=0^{+}$.

Now, the surface energy-momentum tensor of the disk is computed from $Q_{\alpha\beta}$ through the integral
\begin{equation}
S_{\alpha\beta}=\int Q_{\alpha\beta}\delta(z)ds_{n},
\label{eq19}
\end{equation}
where $ds_{n}=\sqrt{g_{zz}}dz$. This integral yield the non-null component of $S_{tt}$, which we write in the tetrad (\ref{E0116}) to interpret it physically as the surface energy density of the disk-like source,
\begin{equation}
\sigma=4e^{\psi}\psi_{,z}|_{_{z=0^{+}}}.
\label{eq20}
\end{equation}
This solution can be rewritten in terms of $U$ as
\begin{equation}
\sigma=\frac{4}{k}\frac{U_{,z}}{\left(1-U\right)^{\frac{k+1}{k}}},
\label{eq20s}
\end{equation}
which is evaluated in $z=0^{+}$

As in the previous section, $\sigma$ needs also fulfill the energy conditions, but in this case it is easier than before because our disk is made of dust ($p=0$), i.e. the particles in the fluid are collisionless; so all the energy conditions are satisfy if
\begin{equation}
\psi_{,z}|_{_{z=0^{+}}}>0.
\label{eq21}
\end{equation}
Here we do not consider the case $\psi_{,z}|_{_{z=0^{+}}}=0$ because it implies $b_{\alpha\beta}=0$ and $\sigma=0$.

\section{\label{sec:mass}Gravitational Mass and Circular Geodesics}

Once the energy-momentum tensor has been specified, it is possible to compute the total mass of the system. To do it, we use the Komar formulae \cite{9}, which is written as
\begin{equation}
M=2\int\left( T_{\alpha\beta}-\frac{1}{2}Tg_{\alpha\beta}\right)m^{\alpha}\xi^{\beta}_{(t)}\sqrt{h}d^{3}y,
\label{eq22}
\end{equation}
where $m_{\alpha}$ is the normal vector to the space-like hypersurface, $\xi^{\beta}_{(t)}$ is the time-like killing vector, $h$ is the determinant of the induced metric and $d^{3}y$ is the volume element. In our case, $m^{\alpha}=e^{-\psi}\delta_{0}^{\alpha}$, $\xi^{\beta}_{(t)}=\delta_{0}^{\beta}$, and $d^{3}y=drd\varphi dz$.

By using the non null components of $T_{\alpha\beta}$ and the relation (\ref{eq10}) in the Komar formulae we can compute the mass of the halo as
\begin{equation}
M_{H}=\frac{2}{k}\int_{V}\frac{\nabla U\cdot\nabla U}{\left(1-U\right)^{2}} dV,
\label{Mh}
\end{equation}
where $dV$ is the usual euclidean volume element of the space. On the other hand, if we take $T_{\alpha\beta}=Q_{\alpha\beta}\delta(z)=e^{\psi}S_{\alpha\beta}\delta(z)$ in the Komar formulae we find that the mass of the disk is
\begin{equation}
M_{D}=\frac{8\pi}{k}\int_{0}^{\infty}\frac{U_{,z}}{1-U}\Bigr|_{z=0^{+}}rdr.
\label{Md}
\end{equation}
Finially, the total mass of the system is given by $M_{T}=M_{D}+M_{H}$.

Now, we can also study the movement of a particle in a spacetime described by the line element (\ref{eq1}). The particle is interacting with a gravitational field produced by a disk-like distribution of energy (\ref{eq20s}) surrounded by a halo whose energy distribution is described by (\ref{eq10u}). Now, a very simple case is the one of a particle that moves in a circular orbit in the plane of the disk ($z=0$), with velocity $v_{c}$ tangential to its path. This is interesting, from the physical point of view, because we have observational information for $v_{c}(r)$ from spiral galaxies \cite{26}. The relation between the tangential velocity and the distant from the center of the galaxy ($r$) is described by the rotational curves.

For a circular path in the plane of the disk $z=0$ and $r$ is constant; therefore, the velocity is described only by two components,
\begin{equation}
u^{\alpha}=u^{0}(1,0,\omega,0),
\label{eq30}
\end{equation}
where $w=u^{2}/u^{0}$ is the angular velocity. By imposing the normalization condition $g_{\mu\nu}u^{\nu}u^{\nu}=-1$, we find that
\begin{equation}
(u^{0})^{2}=\frac{e^{-2\psi}}{1-v_{c}^{2}},
\label{eq31}
\end{equation}
where $v_{c}^{2}=e^{-4\psi} r^{2}\omega^{2}$ is the circular velocity or tangential velocity. On the other hand, we solve the geodesic equation,
\begin{equation}
\frac{du_{\alpha}}{d\tau}=\frac{1}{2}g_{\mu\nu,\alpha}u^{\mu}u^{\nu},
\label{eq32}
\end{equation}
to obtain an expression for $v_{c}$ in terms of the function $U$. As we saw, the spacetime has time symmetry and axial symmetry, so $u^{0}$ and $u^{2}$ are not going to change in $\tau$, so, the geodesic equation reduces to $g_{\mu\nu,r}u^{\mu}u^{\nu}=0$, and it yield the expression for the tangential velocity,
\begin{equation}
v_{c_{n}}^{2}=\frac{rU_{n_{,r}}}{k(1-U_{n})-rU_{n_{,r}}},
\label{eq33}
\end{equation}
from which we obtain the rotational curves.

\section{\label{sec:kuzmin}Kuzmin-Toomre Relativistic Disks with Halo}

The energy density of the halo and the disk, the stress of the halo, the gravitational mass and the circular velocity are the physical properties that describe the disk-halo system. All these properties are written in terms of the solution to Laplace equation $U$. Nevertheless, as we saw in Sec. \ref{sec:emtensor}, $U(r,z)$ needs to satisfy the conditions (\ref{eq15}) and (\ref{eq17u}) in order to define a surface energy density in the plane $z=0$.

To find a particular family of models we choose $U$ to be the solution to Laplace equation in spherical coordinates $(R,\theta,\phi)$. Now, if we use the usual coordinates transformations $R^{2}=r^2+z^2$ and $\cos\theta=z/R$, then the condition (\ref{eq17u}) will not be satisfied because $b_{\alpha\beta}=0$. To solve this problem we use the displacement, cut and reflect method used by Kuzmin (1956) and Toomre (1963) \cite{4,5} to generate disk-like sources in the frame of newtonian gravity. The method is based on the transformation
\begin{equation}
z\rightarrow|z|+a,
\label{kt}
\end{equation}
where $a$ is a constant. With this, the function $U$ takes the form
\begin{equation}
U_{n}(r,z)=-\sum_{l=0}^{n}\frac{A_{l}}{R^{l+1}}P_{l}(\cos\theta),
\label{ukuz}
\end{equation}
where $R^{2}=r^{2}+(|z|+a)^{2}$, $\cos\theta=(|z|+a)/R$, $A_{l}$ are arbitrary constants and $P(\cos\theta)$ are the Legendre Polynomials.

The solution to the Laplace equation presented in (\ref{ukuz}) allows us to write the energy density of the halo as
\begin{widetext}
\begin{equation}
\rho_{n}(r,z) = \frac{2k-1}{k^{2}} \left\{ \frac{\left[\sum_{l=0}^{n}A_{l}rP'_{l+1}(\frac{|z|+a}{R})\right]^{2} + \left[\sum_{l=0}^{n}\left(A_{l}(l+1)P_{l+1}(\frac{|z|+a}{R})\right)/R^{l+2}\right]^{2}}{\left[1+\sum_{l=0}^{n}A_{l}P_{l}(\frac{a}{R})/R^{l+1}\right]^{2(k+1)/k}} \right\}.
\label{rhokuzmin}
\end{equation}
\end{widetext}
In a similar way, we can write the energy density of the disk as
\begin{equation}
\sigma_{n}(r)=\frac{4}{k}\frac{\sum_{l=0}^{n}\left[A_{l}(l+1)P_{l+1}(a/R_{0})\right]/R_{0}^{l+2}}{\left[1+\sum_{l=0}^{n}A_{l}P_{l}(a/R_{0})/R_{0}^{l+1}\right]^{(k+1)/k}},
\label{sigmakuzmin}
\end{equation}
where $R_{0}^{2}=r^{2}+a^{2}$. So, from the behavior of $P(\cos\theta)$ and the fact that $U_{n}(r,z)$ vanishes at infinity, we can see that $\sigma_{n}(r)$ and $\rho_{n}(r,z)$ are always positive, have their maximum value at the center of the system ($r=0$, $z=0$) and both vanish at infinity.

Now, we are going to analyze the behavior of $\rho_{n}$ and $\sigma_{n}$, and to do so we will consider only the models with $n=0,1,2$. The solutions that describe the energy density of the halo for the first three models are given by

\begin{widetext}
\begin{eqnarray}
\tilde{\rho}_{0}&=&\frac{\tilde{A}_{0}^{2}\left( \tilde{A}_{0}+\tilde{R}\right)^{-2(k+1)/k}}{\tilde{R}^{2(k-1)/k}},\label{r0}\\ \nonumber\\
\tilde{\rho}_{1}&=&\frac{\tilde{r}^{2}[\tilde{A}_{0}\tilde{R}^{2}+3\tilde{A}_{1}\tilde{Z}]^{2}+[\tilde{Z}(\tilde{A}_{0}\tilde{R}^{2}+3\tilde{A}_{1}\tilde{Z})-\tilde{A}_{1}\tilde{R}^{2}]^2}{\tilde{R}^{2(2k-3)/k}[\tilde{R}^{3}+\tilde{R}^{2}\tilde{A}_{0}+\tilde{A}_{1}\tilde{Z}]^{2(1+k)/k}}, \label{r1}\\ \nonumber\\
\tilde{\rho}_{2}&=&\frac{\tilde{R}^{2(5-2k)/k}\left[\tilde{A}_{0}\tilde{r}\tilde{R}^{4}+(3\tilde{A}_{1}\tilde{Z}+\tilde{A}_{2})\tilde{r}\tilde{R}^{2}+\frac{5}{2}\tilde{A}_{2}(2\tilde{Z}^{2}-\tilde{r}^{2})\right]^{2}}{\left[\tilde{R}^{5}+\tilde{A}_{0}\tilde{R}^{4}+\tilde{A}_{1}\tilde{R}^{2}\tilde{Z}+\frac{1}{2}\tilde{A}_{2}(\tilde{Z}^{2}-\tilde{r}^{2})\right]^{2(1+k)/k}} \nonumber \\ \nonumber \\
&&+ \frac{\tilde{R}^{2(5-2k)/k}\left[(\tilde{A}_{0}\tilde{Z}-\tilde{A}_{1})\tilde{R}^{4}+3\tilde{Z}(\tilde{A}_{1}\tilde{Z}-\tilde{A}_{2})\tilde{R}^{2}+\frac{5}{2}\tilde{A}_{2}\tilde{Z}(2\tilde{Z}^{2}-\tilde{r}^{2})\right]^{2}}{\left[\tilde{R}^{5}+\tilde{A}_{0}\tilde{R}^{4}+\tilde{A}_{1}\tilde{R}^{2}\tilde{Z}+\frac{1}{2}\tilde{A}_{2}(\tilde{Z}^{2}-\tilde{r}^{2})\right]^{2(1+k)/k}}, \label{r2}
\end{eqnarray} \\
where $\tilde{R}^{2}=\tilde{r}^{2}+\tilde{Z}^{2}$, $\tilde{Z}=|\tilde{z}|+1$, $\tilde{z}=z/a$, $\tilde{r}=r/a$, $\tilde{A}_{0}=A_{0}/a$, $\tilde{A}_{1}=A_{1}/a^{2}$, $\tilde{A}_{2}=A_{2}/a^{3}$, and $\tilde{\rho}_{n}=[(ka)^{2}/(2k-1)]\rho_{n}$.
\end{widetext}

\begin{figure}
$$\begin{array}{c}
n = 0 \\
\epsfig{width=2.3in,file=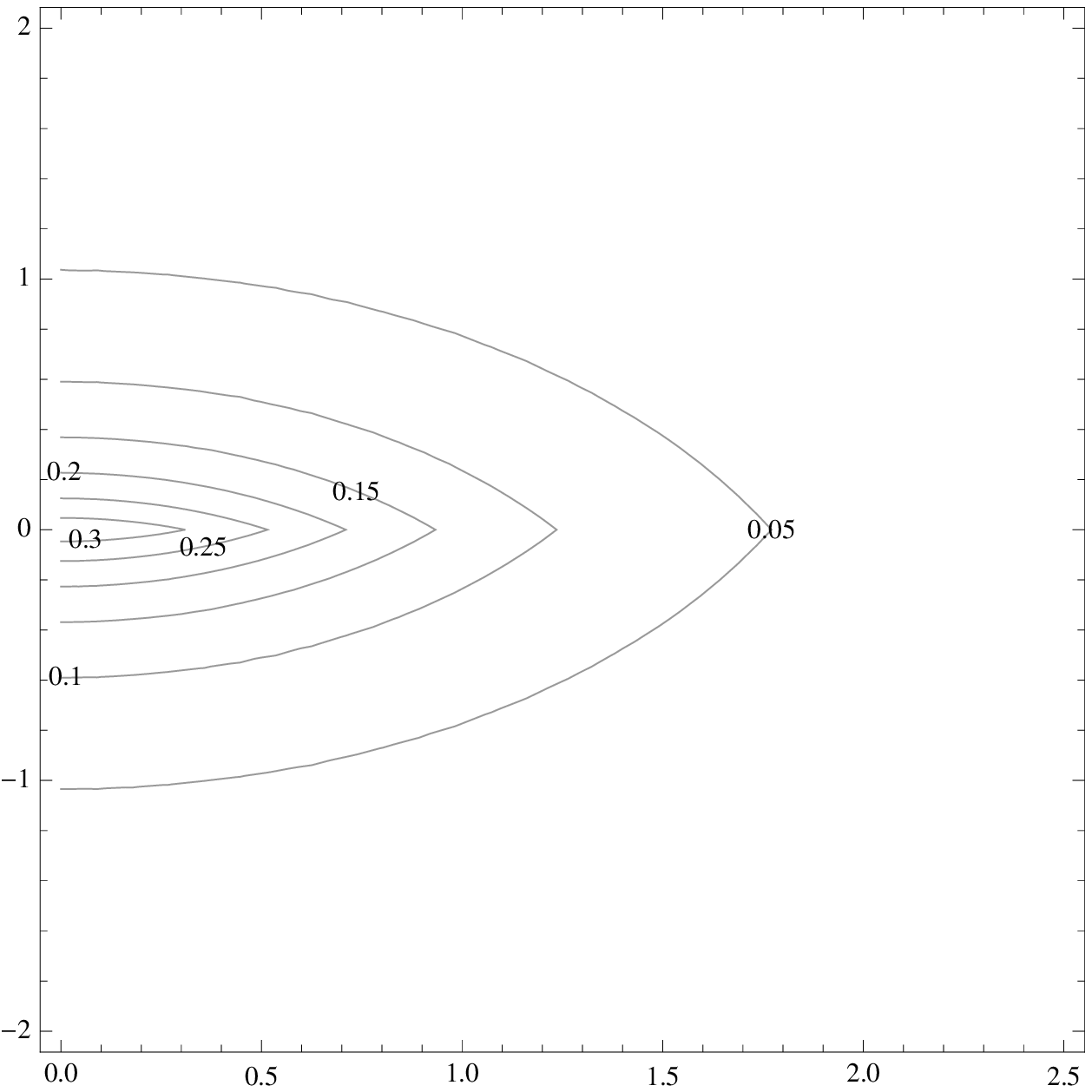} \\ \\
n = 1 \\
\epsfig{width=2.3in,file=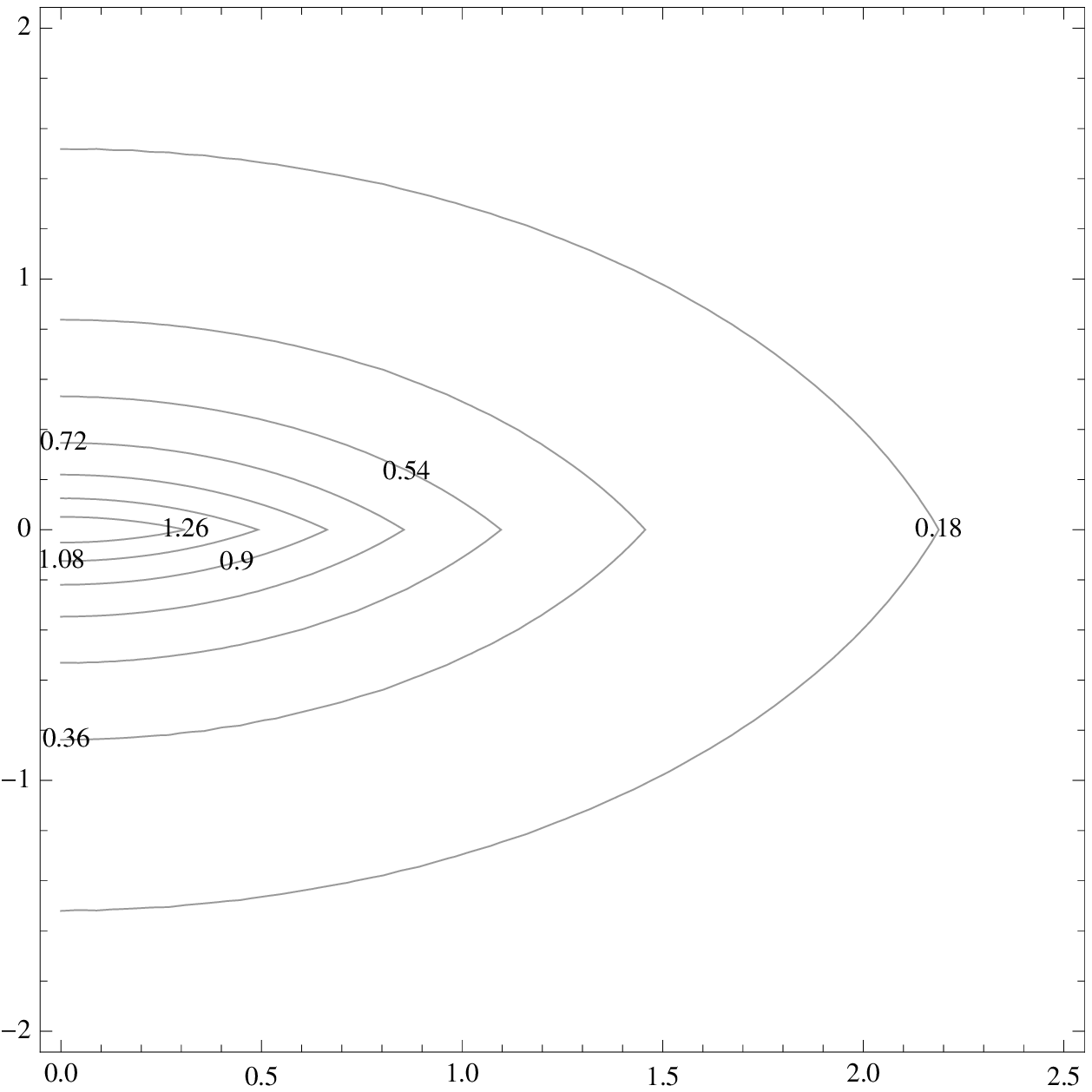} \\ \\
n = 2 \\
\epsfig{width=2.3in,file=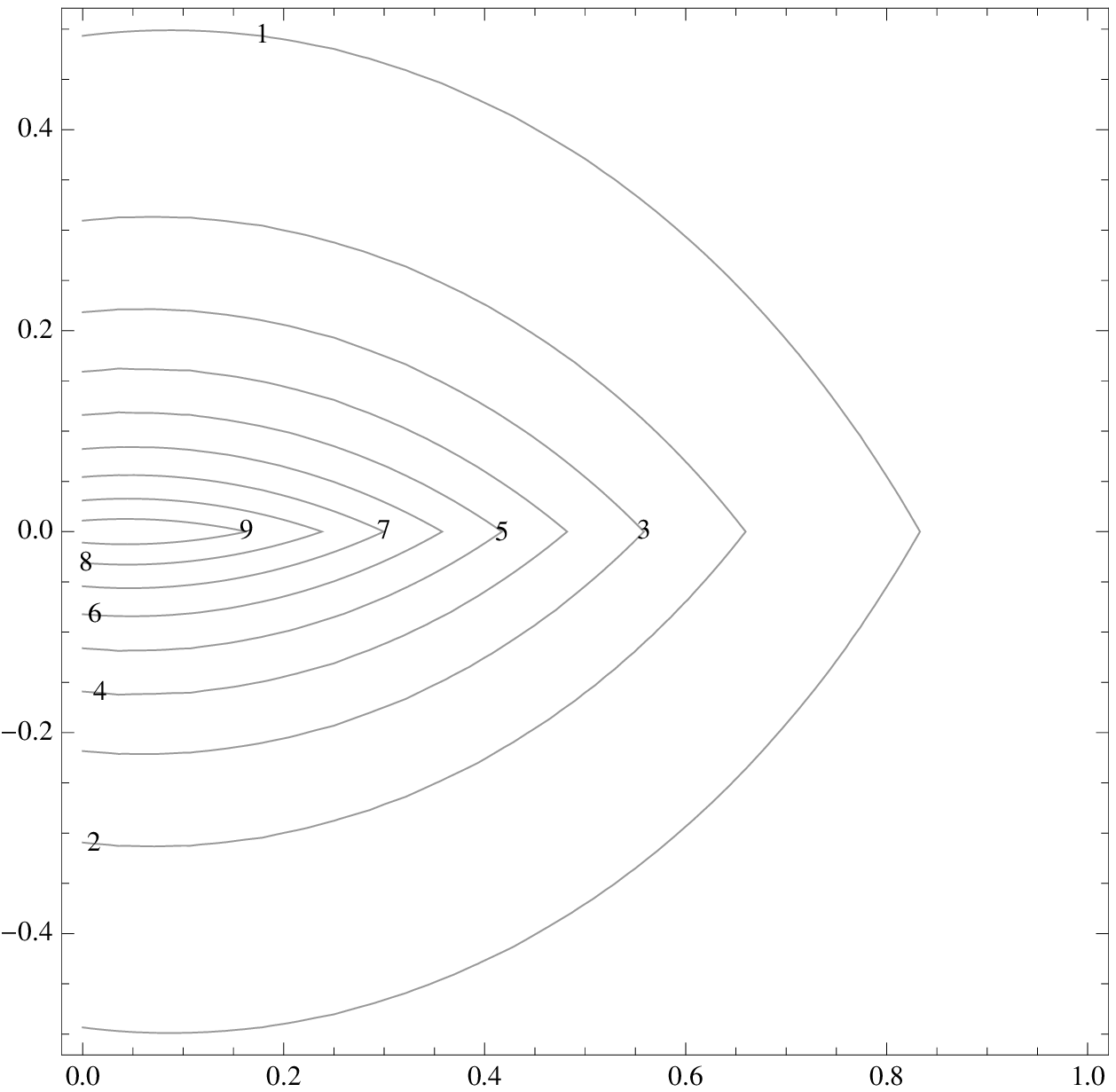} \\
\end{array}$$
\caption{Contour plots of energy density of the halo $\tilde{\rho}_{n}$ as a function of $\tilde{r}$ and $\tilde{z}$ for the first 3 models of the family of solutions. The contours for the $n=0$ model correspond to the values $k=7.9, \tilde{A}_{0}=2$; the contours for the $n=1$ model to $k=292, \tilde{A}_{0}=167,$ and $\tilde{A}_{1}=46$; and those for $n=2$ model correspond to the values $k=18.12, \tilde{A}_{0}=100, \tilde{A}_{1}=-20$ and $\tilde{A}_{2}=90$. In each plot, the vertical axis corresponds to $\tilde{z}$ coordinate, and the horizontal one to $\tilde{r}$ coordinate.\label{denhalokuz}}
\end{figure}

In FIG. \ref{denhalokuz} we plot (\ref{r0}-\ref{r2}) to analyze the behavior of the energy density of the halo. This figure allows us to see that the most of the energy in the halo is located at the center of the system, i.e. in the region near to $\tilde{r}=0$ and  $\tilde{z}=0$. The density presents a maximum at the center and, it tends to zero at infinity with a rate that depends on the constants values in each particular model.

On the other hand, the energy density of the disk for the models $n=0,1,2$ is given by
\begin{widetext}
\begin{eqnarray}
\tilde{\sigma}_{0}&=&\frac{\tilde{A}_{0}\left(\tilde{A}_{0}+\tilde{R}_{0}\right)^{-\frac{1+k}{k}}}{\left(\tilde{R}_{0}\right)^{\frac{2k-1}{k}}},\label{s0}\\ \nonumber \\
\tilde{\sigma}_{1}&=&\frac{\left[ \tilde{A}_{0}\tilde{R}_{0}^{2}+\tilde{A}_{1}(2-\tilde{r}^2)\right]\tilde{R}_{0}^{\frac{3-2k}{k}}}{\left[\tilde{R}_{0}^{3}+\tilde{A}_{0}\tilde{R}_{0}^{2}+\tilde{A}_{1}\right]^{\frac{1+k}{k}}},\label{s1}\\ \nonumber \\
\tilde{\sigma}_{2}&=&\frac{\left[(\tilde{A}_{0}-\tilde{A}_{1})\tilde{R}_{0}^{4}+3(\tilde{A}_{1}-\tilde{A}_{2})\tilde{R}_{0}^{2}+\frac{5}{2}\tilde{A}_{2}(2-\tilde{r}^2)\right]\tilde{R}_{0}^{\frac{2(5-2k)}{k}}}{\left[2\tilde{R}_{0}^{5}+2\tilde{A}_{0}\tilde{R}_{0}^{4}+2\tilde{A}_{1}\tilde{R}_{0}^{2}+\tilde{A}_{2}(2-\tilde{r}^2)\right]^{\frac{1+k}{k}}},\label{s2}
\end{eqnarray}
where $\tilde{R}_{0}^{2}=1+\tilde{r}^{2}$, $\tilde{r}=r/a$, $\tilde{A}_{0}=A_{0}/a$, $\tilde{A}_{1}=A_{1}/a^{2}$, $\tilde{A}_{2}=A_{2}/a^{3}$, and $\tilde{\sigma}_{n}=(ka/4)\sigma_{n}$.
\end{widetext}

\begin{figure}
$$\begin{array}{c}
n = 0 \\
\epsfig{width=3.2in,file=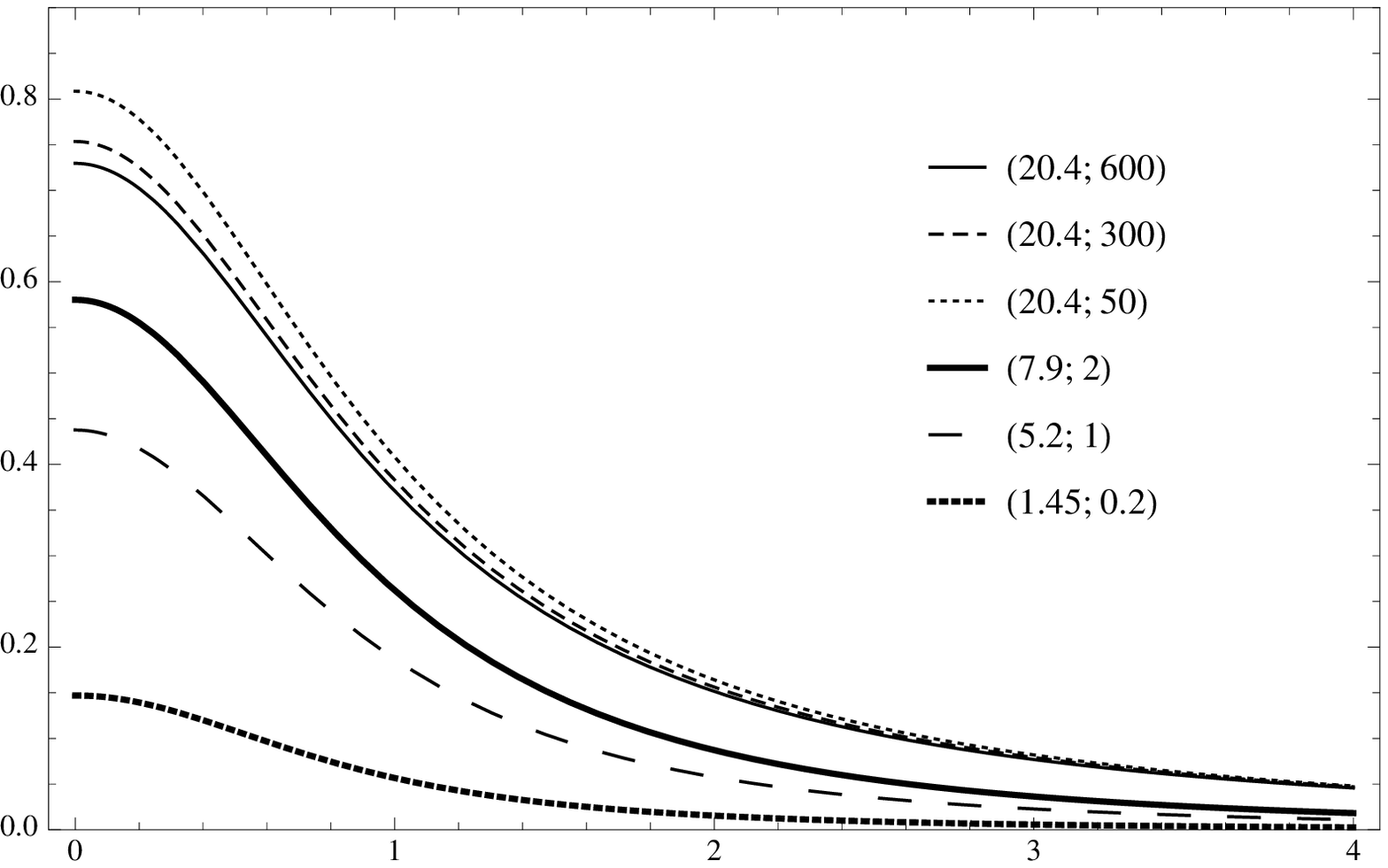} \\ \\
n = 1 \\
\epsfig{width=3.2in,file=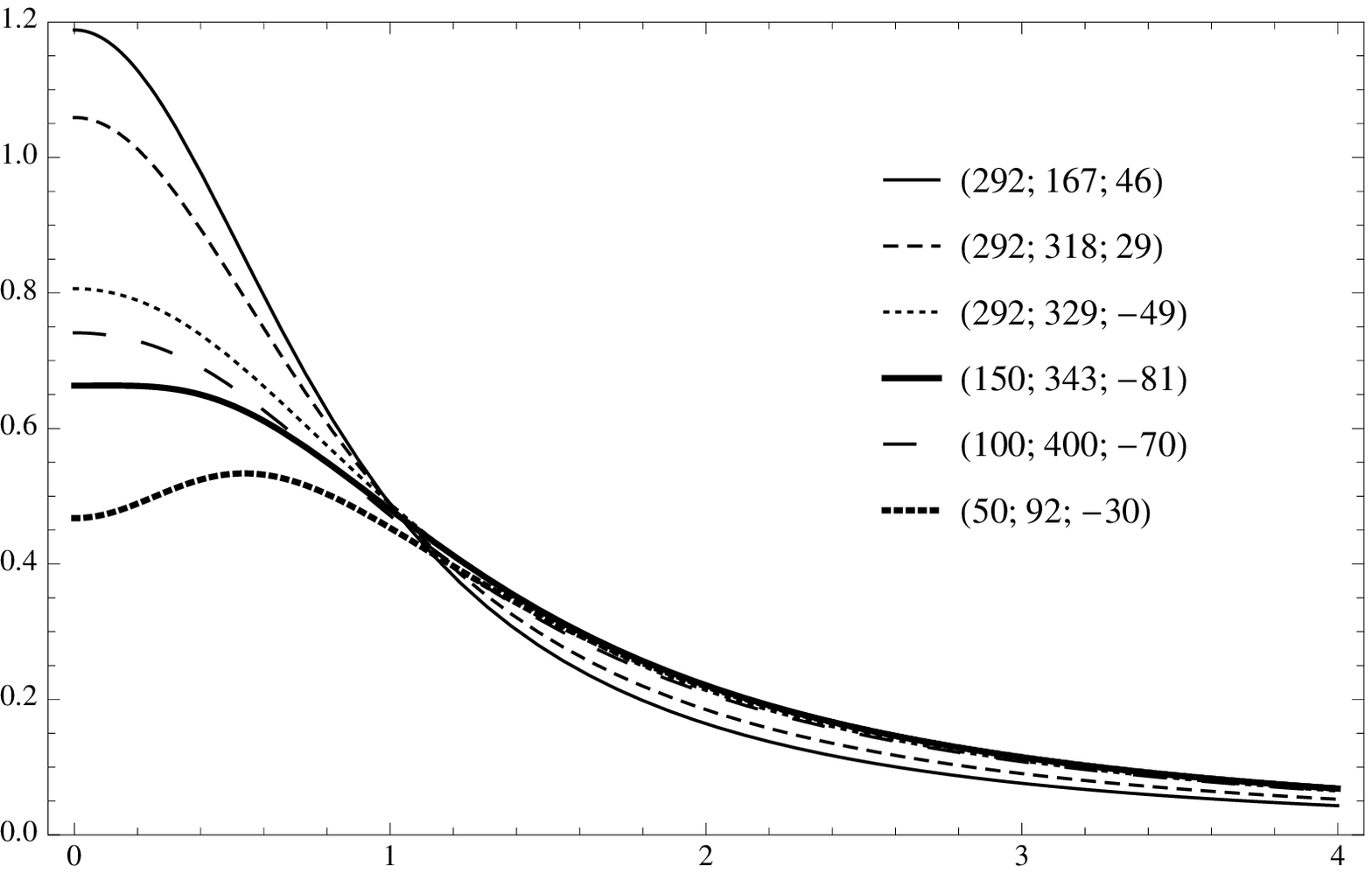} \\ \\
n = 2 \\
\epsfig{width=3.2in,file=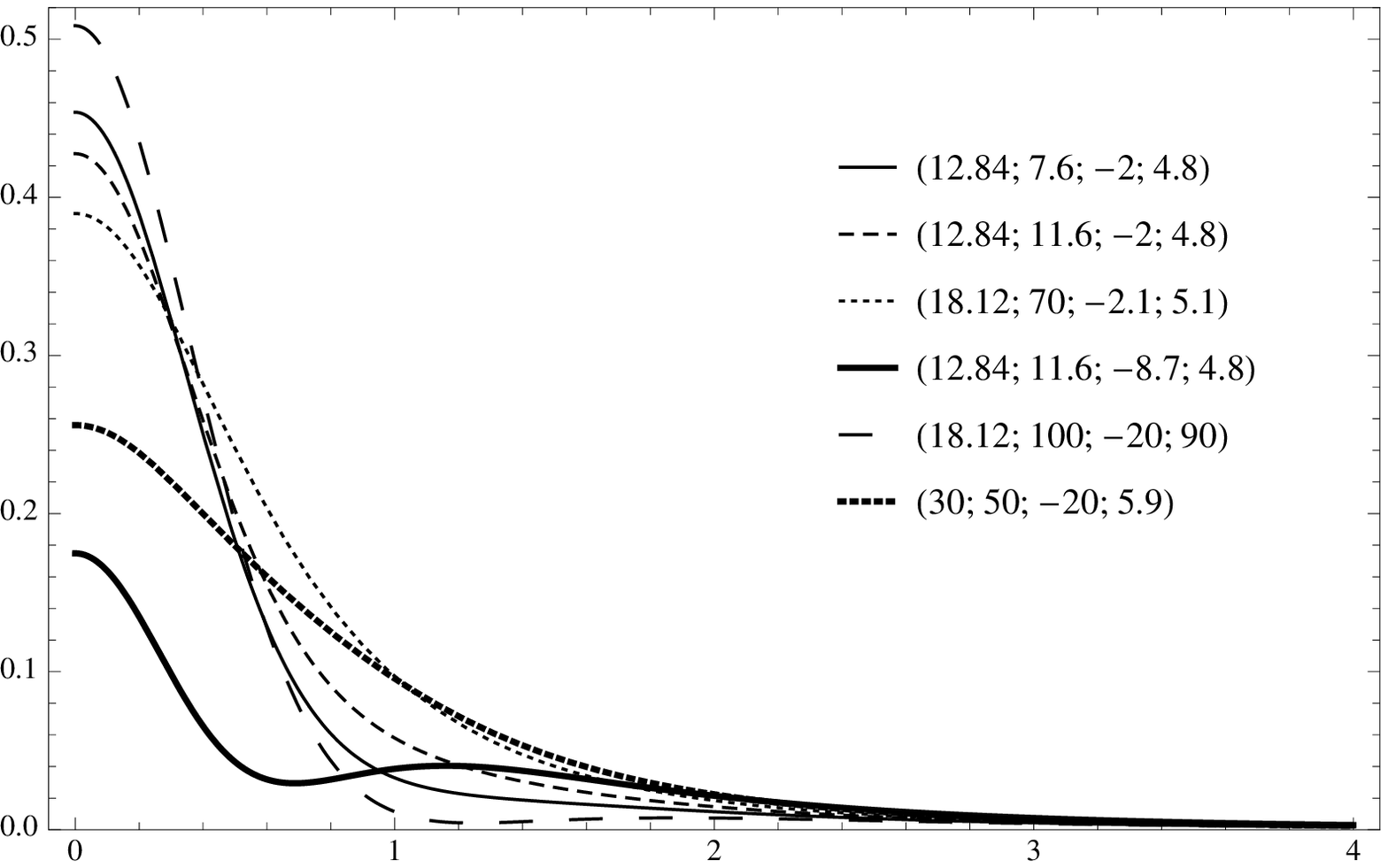} \\
\end{array}$$
\caption{\label{densidiskuz}Plots of the surface energy density $\tilde{\sigma}_{n}$ of the disk as a function of $\tilde{r}$ for the first $3$ models of the family of solutions. Each curve is labeled by a set of numbers so that, for the $n$-model the numbers are ($k, \tilde{A_{0}}, \tilde{A_{1}}, \tilde{A_{2}},..., \tilde{A_{n}}$).}
\end{figure}

We plot, in FIG. \ref{densidiskuz}, the energy density $\tilde{\sigma}$ for the first models with $n=1,2,3$. In each graph we plot some curves for different values of the constants $\tilde{A}$ and $k$ in order to analyze different behaviors. Hereafter, we introduce a convenient notation in which each curve is labeled by the set of numbers ($k, \tilde{A_{0}}, \tilde{A_{1}}, \tilde{A_{2}},..., \tilde{A_{n}}$). It is important to clarify that the constants are chosen so that the energy density is positive; this assure the condition (\ref{eq21}).

The surface energy density profiles, presented in FIG. \ref{densidiskuz}, go to zero at infinity from a maximum value located at the center of the disk. We can change the maximum value of $\tilde{\sigma}_{n}$ and its decrease rate by changing the values of the constants. In the three models, the most of the energy is concentrated at the central region of the disk, but as we can see, when we increase $n$, the energy is less distributed in the space, i.e. the energy density goes to zero faster.

We can compute the total mass for the disk-halo system, which is defined through (\ref{Mh}) and (\ref{Md}), for any model characterized by the integer value of $n$. By doing the integrals for $n=0$ we find that,
\begin{eqnarray}
M_{T_{0}}&=&M_{D}+M_{H}=\frac{8\pi A_{0}}{k},\label{M}\\
M_{D_{0}}&=&\frac{8\pi a}{k}\ln\left(1+\frac{A_{0}}{k}\right),\label{Md0}\\
M_{H_{0}}&=&\frac{8\pi a}{k}\left[ \frac{A_{0}}{a}-\ln\left(1+\frac{A_{0}}{k}\right)\right].\label{Mh0}
\end{eqnarray}
Now, since $\sigma_{n}$ and $\rho_{n}$ vanish at infinity, it is important to show that $M_{T}$ will converge for all $n$. To do so, we start by putting (\ref{Md}) in the form
\begin{equation}
M_{D_{n}}=\frac{8\pi}{k}\int_{a}^{\infty}\mu_{n}(R_{0})R_{0}dR_{0},
\label{mdproof}
\end{equation}
where $R_{0}^{2}=r^{2}+a^{2}$ and
\begin{equation}
\mu_{n}(R_{0}) = \frac{U_{n_{,z}}}{1-U}\Bigr|_{z=0^{+}}.
\end{equation}
With (\ref{ukuz}) it is easy to show that 
\begin{equation}
\lim_{R_{0}\rightarrow\infty}\frac{\mu_{n+1}}{\mu_{n}}=1;
\label{cp}
\end{equation}
and by using the limit comparison test \cite{10} we can demonstrate that $M_{D_{n+1}}$ will be finite if $M_{D_{n}}$ converges. Now, since $M_{D_{0}}$ has a finite value, then the mass of the disk for every model will be finite too.

To guarantee the convergence of $M_{H_{n}}$ we proceed in a similar manner: the integral (\ref{Mh}), with 
\begin{equation}
dV = r^{2} \sin \theta dr d\theta d\varphi,
\end{equation}
can be rewritten as
\begin{equation}
M_{H_{n}}=\frac{4\pi}{k}\int_{-1}^{1}\int_{0}^{\infty}\eta_{n}(R,\tau)RdRd\tau,
\label{mhproof}
\end{equation}
where
\begin{equation}
\eta_{n}(R,\tau) = \frac{\nabla U_{n}\cdot\nabla U_{n}}{(1-U_{n})^{2}}
\end{equation}
and $\tau=\cos\theta$. It is easy to show that
\begin{equation}
\lim_{R\to +\infty}\frac{\eta_{n+1}}{\eta_{n}}=1,
\label{cphalo}
\end{equation}
so by virtue of the limit comparison test, if
$$
F_{n}(\tau)=\int_{0}^{\infty}\eta_{n}(R,\tau)RdR
$$
converge, then $F_{n+1}(\tau)$ will converge too and from (\ref{mhproof}), $M_{H_{n}}$ and $M_{H_{n+1}}$ will be finite. Since $M_{H_{0}}$ has a finite value, then the mass of the halo for all models converge.

Finally, we present the circular velocity for the models we are considering. To compute the expressions for this physical quantity we use (\ref{ukuz}) and (\ref{eq33}) to obtain

\begin{widetext}
\begin{eqnarray}
v_{c_{0}}^{2}&=&\frac{\tilde{A}_{0}\tilde{r}^{2}}{k(\sqrt{m}+\tilde{A}_{0})m-\tilde{A}_{0}\tilde{r}^{2}},\label{v0}\\ \nonumber \\ 
v_{c_{1}}^{2}&=&\frac{\tilde{r}^{2}(\tilde{A}_{0}m+3\tilde{A}_{1})}{k(m^{3/2}+\tilde{A}_{0}m+\tilde{A}_{1})m-\tilde{r}^{2}(\tilde{A}_{0}m+3\tilde{A}_{1})}, \label{v1}\\ \nonumber \\
v_{c_{2}}^{2}&=&\frac{\tilde{r}^{2}[3\tilde{A}_{2}+3\tilde{A}_{1}m^{2}+\tilde{A}_{0}m^{4}+\frac{3}{2}\tilde{A}_{2}(3-m)m]}{k[m^{2}+m\tilde{A}_{1}+m^{3}\tilde{A}_{0}+\frac{1}{2}\tilde{A}_{2}(3-m)]m^{3}-\tilde{r}^{2}[3\tilde{A}_{2}+3\tilde{A}_{1}m^{2}+\tilde{A}_{0}m^{4}+\frac{3}{2}\tilde{A}_{2}(3-m)m]}, \label{v2}
\end{eqnarray} \\
where $m=1+\tilde{r}^{2}$, $\tilde{r}=r/a$, $\tilde{A}_{0}=A_{0}/a$, $\tilde{A}_{1}=A_{1}/a^{2}$, and $\tilde{A}_{2}=A_{2}/a^{3}$.
\end{widetext}

\begin{figure}
$$\begin{array}{c}
n = 0 \\
\epsfig{width=3.2in,file=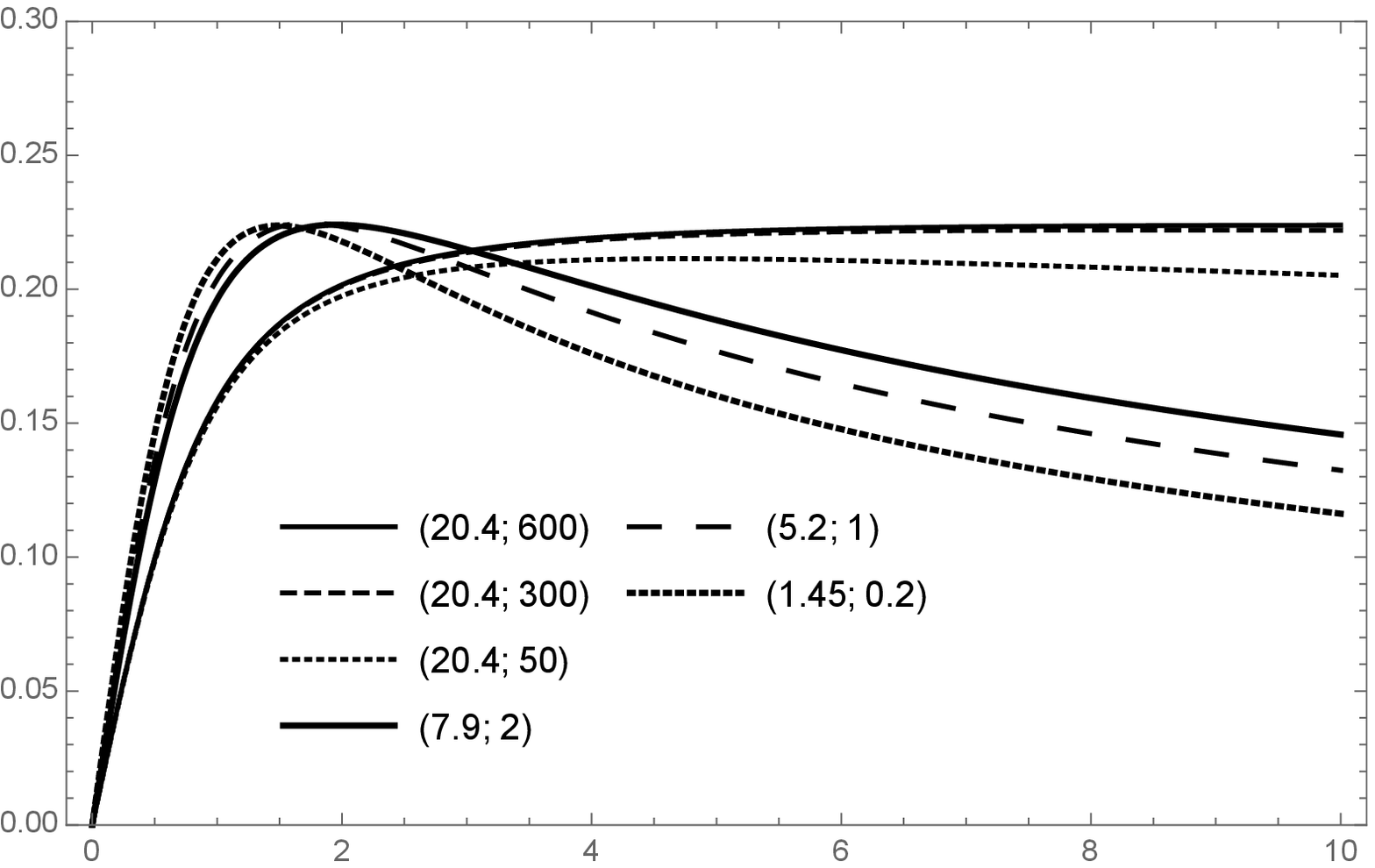} \\ \\
n = 1 \\
\epsfig{width=3.2in,file=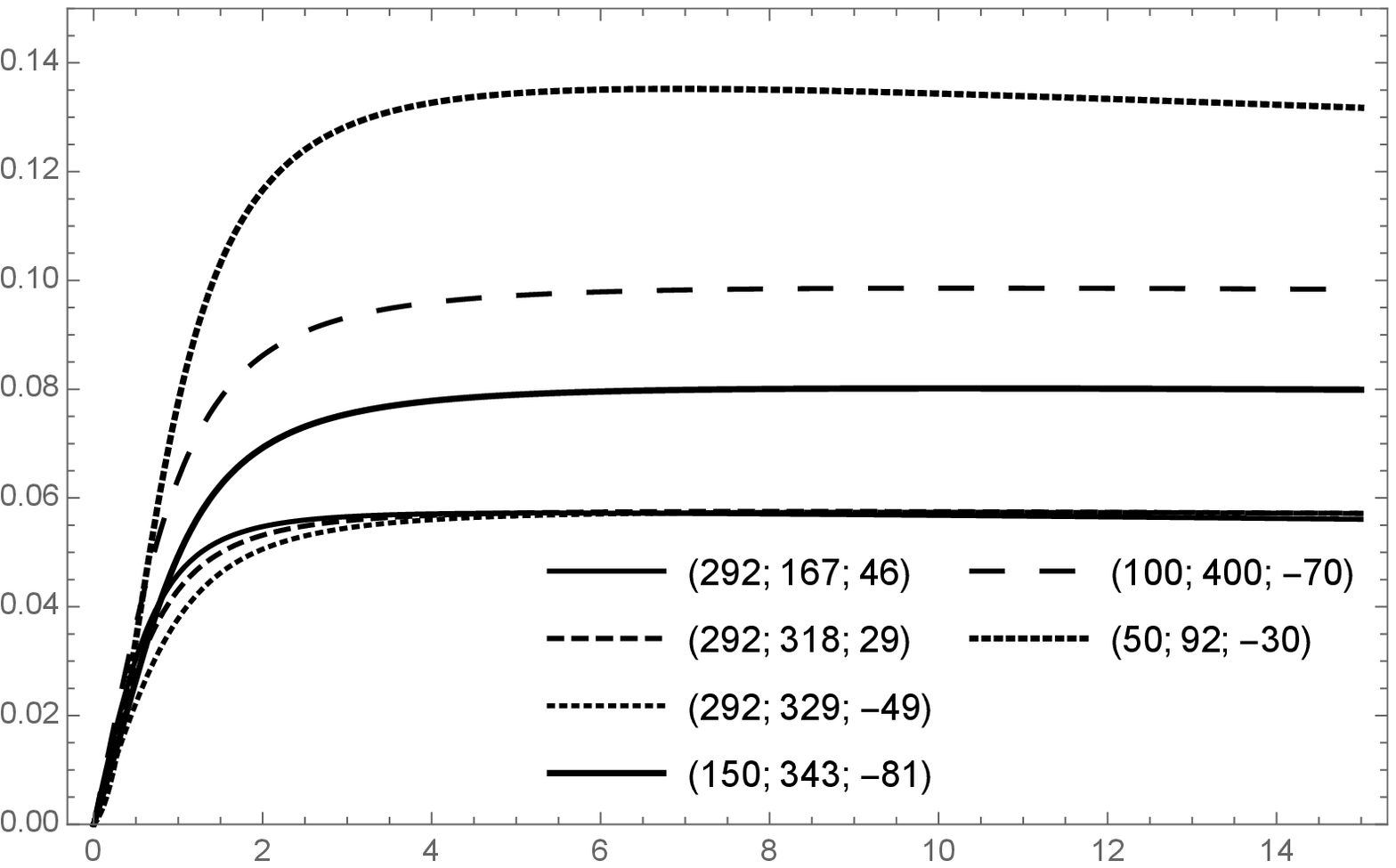} \\ \\
n = 2 \\
\epsfig{width=3.2in,file=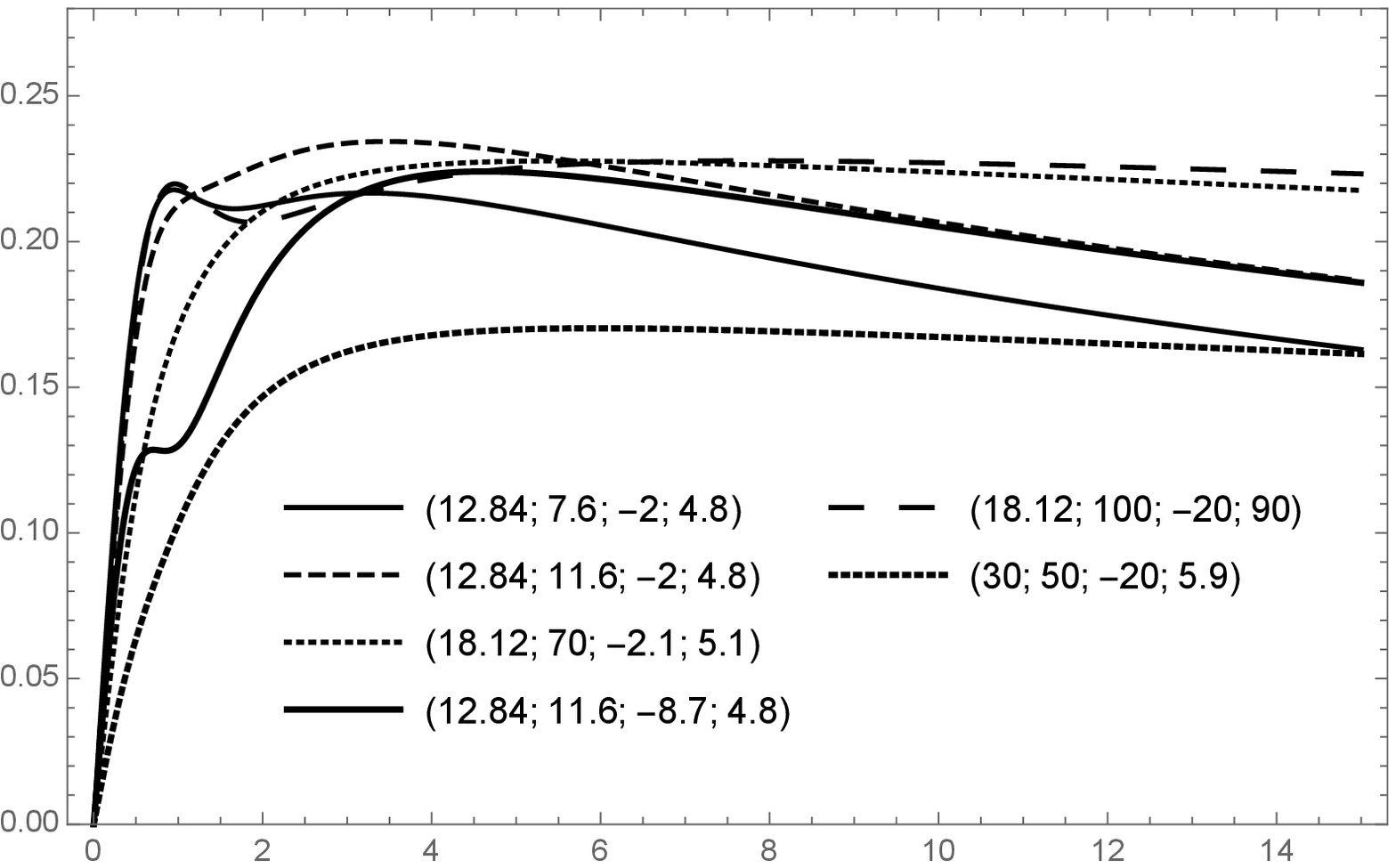} \\
\end{array}$$
\caption{Circular velocity $v_{c_{n}}$ as a function of $\tilde{r}$ for the first 3 models of the family of solutions. Each curve is labeled by a set of numbers so that, for the $n$-model the numbers are ($k, \tilde{A_{0}}, \tilde{A_{1}}, \tilde{A_{2}},..., \tilde{A_{n}}$).\label{circvelkuz}}
\end{figure}

Now, we plot these expressions to see clearly the dependence of $v_{c_{n}}$ with the coordinate $\tilde{r}$ for a time-like particle in a circular orbit. The resulting plots are presented in FIG. \ref{circvelkuz}. Here we choose the same constants values that we use to perform FIG. \ref{densidiskuz} in order to compare the rotational curve with its corresponding surface energy density profile.

In the plot for $n=0$ model we see that the velocity increases very fast until a maximum value from which we distinguish clearly two kind of behavior: in the first one, the velocity decrease  with $\tilde{r}$, and in the second one it remains approximately constant. If we compare with the corresponding plot in FIG. \ref{densidiskuz}, we see that the second kind of behavior is associated with energy density profiles whose maximum is higher than the cases where the velocity decrease after reaching its maximum. 

In the plot for $n=1$, all the curves show that the velocity remains constant after reaching a maximum, but, by comparing with FIG. \ref{densidiskuz}, we see that if the maximum is higher, the energy density reaches a lower value in $\tilde{r}=0$. In the plot for $v_{c_{2}}$ we obtain different kind of behavior. Of particular interest are those with two peaks because actually, the observational data shows that the velocity is not exactly constant after the first maximum, it has fluctuations qualitatively similar to those of our plots. Finally, comparing with the plot for $n=2$ in FIG. \ref{densidiskuz}, we see that, when the rotational curves present two peaks, the corresponding energy density profiles have two peaks as well.

\section{\label{sec:morgan}Morgan-Morgan Relativistic Disks With Halo}


In order to obtain a new family of finite thin disks with halo, we use oblate spheroidal coordinates ($\xi,\eta,\varphi$) to solve the Laplace equation, because they adapt to the geometry of the source and introduce, in a natural way, a finite radius for the disk. The transformation relations with the cylindrical coordinates are

\begin{eqnarray}
r^{2} &=& d^{2}(1 + \xi^{2})(1 - \eta^{2}), \label{robla}\\ 
z &=& d \xi \eta,\label{zobla}
\end{eqnarray} 
where $0\leq\xi\leq\infty$ and $-1\leq\eta\leq1$. The coordinates of the disk of radius $d$ are $\xi=0$ and $0\leq\eta^{2}<1$, and if we cross the disk, $\eta$ changes sign.

In these coordinates, the solution to the Laplace equation takes the form
\begin{equation}
U(\xi,\eta)=-\sum_{n=0}^{\infty}C_{2n}q_{2n}(\xi)P_{2n}(\eta),
\label{uobla}
\end{equation} 
where $P_{2n}(\eta)$ are the Legendre polynomials of order $2n$, $q_{2n}=i^{2n+1}Q_{2n}(i\xi)$, with $Q_{2n}(i\xi)$ the Legendre function of second kind of imaginary argument \cite{11}, and $C_{2n}$ are constants choosen to satisfy both the boundary condition (\ref{eq14}) and the condition for the allowed values for the speed of a particle in a circular trajectory, $0\leq v_{c}\leq1$. According to (\ref{eq15}) and (\ref{eq17u}), to define properly a disk-like source, we want the metric tensor to be continuous and its first $z$-derivate to be discontinuous across the disk; therefore, since $\eta^{2}$ is continuous across the disk and $\eta$ is discontinuous, we only need to consider even Legendre polynomials. Therefore, $U$ in (\ref{uobla}) is formed only by the even terms of the general solution to Laplace equation.

To analyze the behavior of the physical quantities we have to consider particular solutions by restricting the sum in (\ref{uobla}). One way to do it is to consider $U$ to be the gravitational newtonian potential of a finite disk, where the constants $C_{2n}$ are found by imposing the orthogonality property of the Legendre polynomials to give \cite{12}
\begin{equation}
C_{2n}=\frac{K_{2n}}{(2n+1)q_{2n+1}(0)},
\label{eq34}
\end{equation}
where,
\begin{equation}
K_{2n}= \frac{M}{2d}\left[ \frac{\pi^{1/2}(4n+1)(2m+1)!}{2^{2m}(m-n)!\Gamma(m+n+3/2)}\right],
\label{eq35}
\end{equation}
if, $n \leq m$, and $C_{2n}=0$ if $n>m$, for $m\geq1$. In the above equation $M$ and $d$ are the total mass and the radius of the disk respectively. Depending of the value of $m$ we will have a particular solution because the constants $C_{2n}$ are restricted by $ n \leq m$; so we are going to choose only the three first models ($m=1,2,3$) to see the behavior of the energy densities of the disk and the halo, their masses, the state equation of the halo and the rotational curves. 

The energy density of the halo for the first three models are computed by using (\ref{uobla}) in (\ref{eq10u}) in such a way that
\begin{equation}
\rho_{m}=\frac{2k-1}{k^{2}} \left[ \frac{U_{m,r}^{2}+U_{m,z}^{2}}{(1-U_{m})^{2(k+1)/k}} \right],
\label{eq35}
\end{equation}
where
\begin{widetext}\begin{subequations}\label{u}
\begin{align}
U_{1} &= - {\tilde M} \left[ \acot \xi  + A (3\eta^{2} - 1) \right], \\
U_{2} &= - {\tilde M} \left[ \acot \xi + \frac{10 A}{7}
(3\eta^{2} - 1) + B ( 35 \eta^{4} - 30 \eta^{2} + 3) \right],  \\
U_{3} &= - {\tilde M} \left[ \acot \xi + \frac{10 A}{6} (3
\eta^{2} - 1)  + \frac{21 B}{11} (35 \eta^{4} - 30 \eta^{2} + 3) 
+ C (231 \eta^{6} - 315 \eta^{4} + 105 \eta^{2} - 5) \right],
\end{align}\end{subequations}
with ${\tilde M} = M/d$ and
\begin{subequations}\begin{align}
A &= \frac{1}{4} \left[ (3\xi^{2} + 1) \acot \xi - 3 \xi \right],
\\
B &= \frac{3}{448} \left[ (35 \xi^{4} + 30 \xi^{2} + 3) \acot \xi
- 35 \xi^{3} - \frac{55}{3} \xi \right], \\
C &= \frac{5}{8448} \left[ (231 \xi^{6} + 315 \xi^{4} + 105 \xi^{2} +
5) \acot \xi  - 231 \xi^{5} - 238 \xi^{3} - \frac{231}{5} \xi \right].
\end{align}
\end{subequations}
\end{widetext}
Note that $\xi$ and $\eta$ are related with the cylindrical coordinates by virtue of the transformation relations (\ref{robla}) and (\ref{zobla}), so we can compute the first $r$-derivate of the above expressions,
\begin{widetext}
\begin{subequations}\label{ur}
\begin{align}
U_{1,r} =& \frac{3\tilde{M}\tilde{r}}{2d(1+\xi^{2})}\left[-\xi+(1+\xi^{2}) \acot \xi \right], \\
U_{2,r} =& \frac{15\tilde{M}\tilde{r}}{16d(1+\xi^{2})} \left\{ \xi-13\eta^{2}\xi+(3-15\eta^{2})\xi^{3}+(1+\xi^{2}) [ 1 - 3\xi^{2}+3\eta^{2}(1+5\xi^{2}) ] \acot \xi\right\}, \\
U_{3,r} =& \frac{35\tilde{M}\tilde{r}}{128d(1+\xi^{2})}\left\{ \xi \left[ 3 - 4 \xi^{2} - 15 \xi^{4} + \eta^{2} (6 + 200 \xi^{2} + 210 \xi^{4}) - \eta^{4}(113 + 420 \xi^{2} + 315 \xi^{4}) \right] \right. \notag \\
&\left. + 3 (1+\xi^{2}) \left[ 1 - 2 \xi^{2} + 5 \xi^{4} + \eta^{2}(2 - \xi^{2}-70\xi^{4})+5\eta^{4}(1+14\xi^{2}+21\xi^{4}) \right] \acot \xi \right\},
\end{align}
\end{subequations}
and also, the first $z$-derivative, which are
\begin{subequations}\label{uz}
\begin{align}
U_{1,z} =& -\frac{3\tilde{M}\eta}{d} \left[ - 1 + \xi \acot \xi \right], \\
U_{2,z} =& \frac{5\tilde{M}\eta}{4d} \left\{ -9 \xi^{2} + \eta^{2} (4 + 15 \xi^{2})+ 3 \xi \left[ 1 + 3 \xi^{2} - \eta^{2} (3 + 5 \xi^{2}) \right] \acot \xi \right\}, \\
U_{3,z} =& -\frac{7\tilde{M}\eta}{64d} \left\{ 50 \eta^{2} \xi^{2}(11 + 21 \xi^{2}) - 15 (\xi^{2} + 15 \xi^{4}) - \eta^{4} (64 + 735 \xi^{2} + 945 \xi^{4}) \right. \notag \\
&\left. + 15 \xi \left[ -1 + 6 \xi^{2} + 15 \xi^{4} - 2 \eta^{2}(3 + 30 \xi^{2} + 35 \xi^{4}) + \eta^{4}(15 + 70 \xi^{2} + 63 \xi^{4}) \right] \acot \xi \right\}.
\end{align}
\end{subequations}
\end{widetext}

\begin{figure}
$$\begin{array}{c}
m = 1 \\
\epsfig{width=2.35in,file=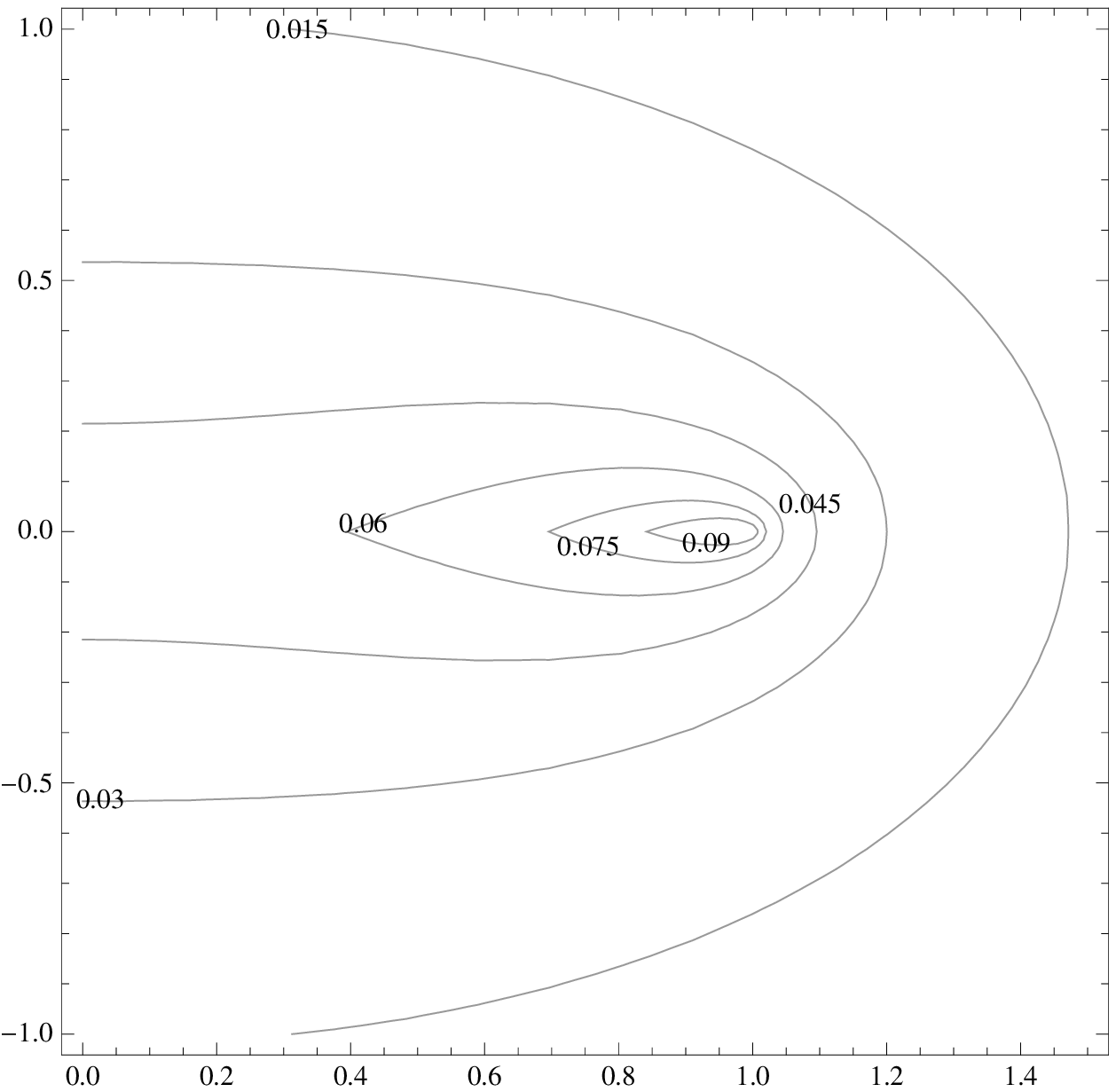} \\ \\
m = 2 \\
\epsfig{width=2.35in,file=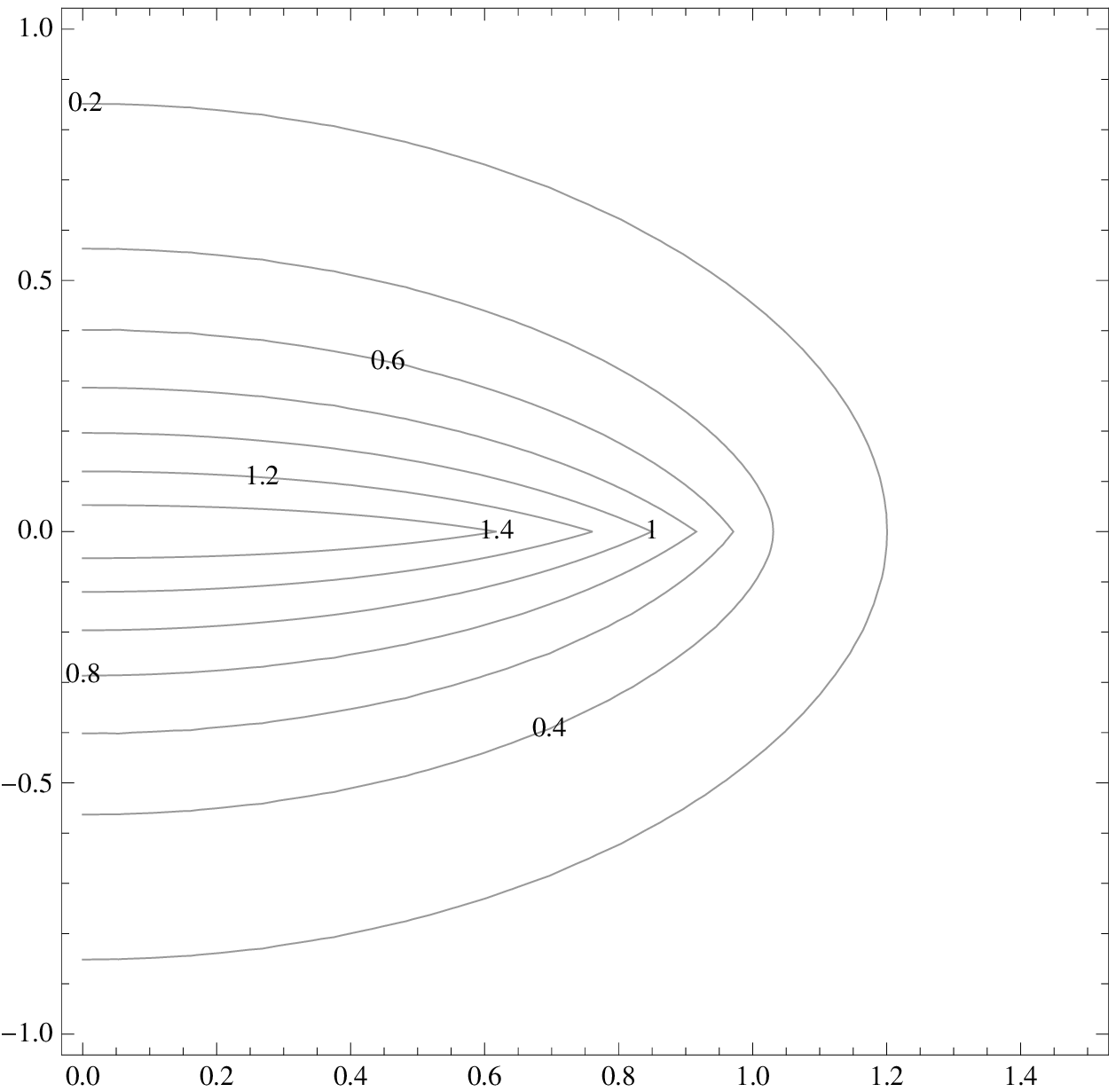} \\ \\
m = 3 \\
\epsfig{width=2.35in,file=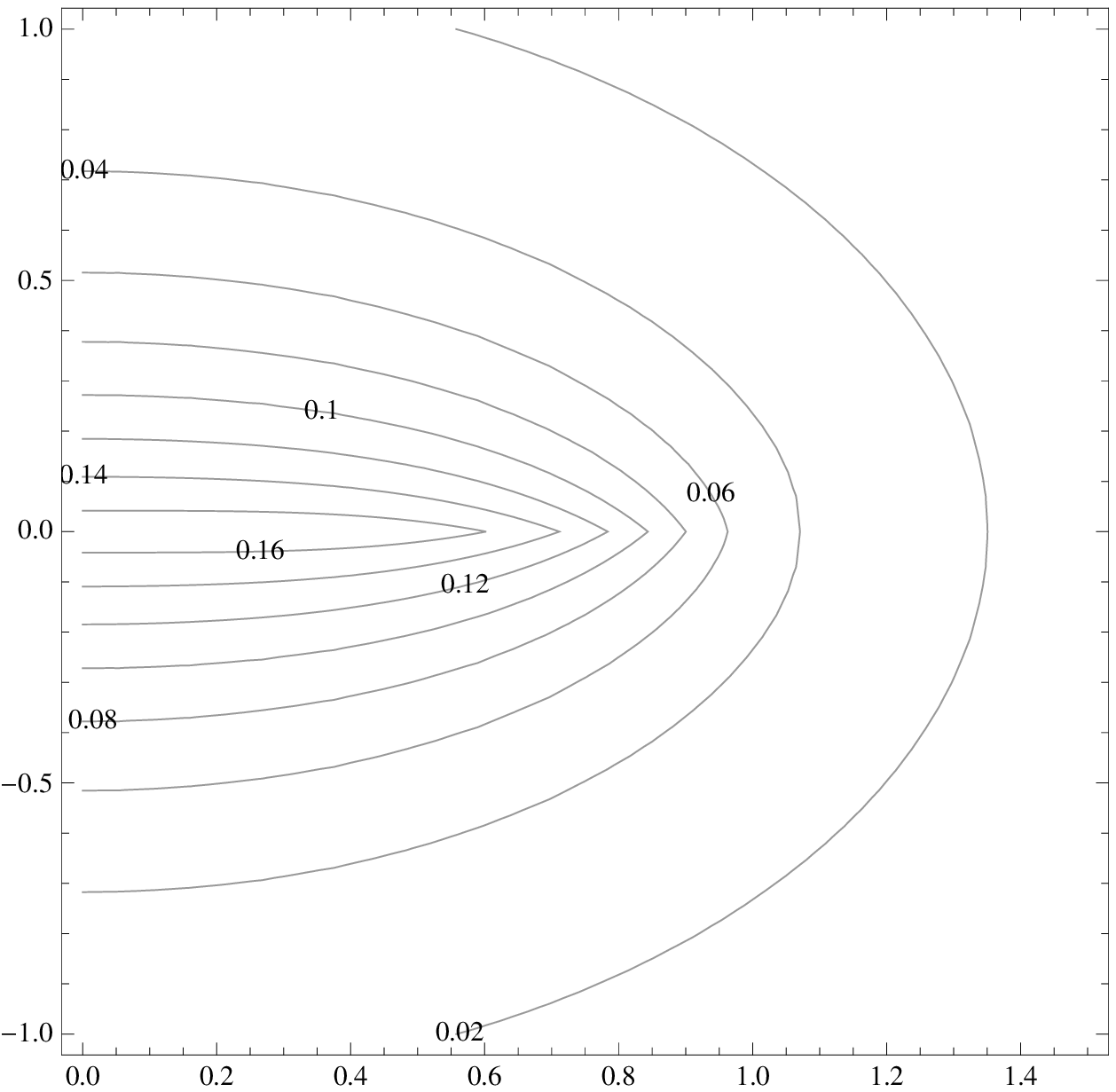} \\
\end{array}$$
\caption{\label{densihalokal}Contour plots of energy density of the halo $\tilde{\rho}_{m}$ as a function of $\tilde{r}$ and $\tilde{z}$ for the first 3 models of the family of solutions. The contours for the $m=1$ model correspond to the values $k=24.5, \tilde{M}=4.5$; the contours for the $m=2$ model to $k=132, \tilde{M}=1$; and those for $m=3$ model correspond to the values $k=17.2, \tilde{M}=3.9$. In each plot, the vertical axis corresponds to $\tilde{z}$ coordinate, and the horizontal one to $\tilde{r}$ coordinate.}
\end{figure} 

In FIG. \ref{densihalokal}, we show the plots of the rescaled energy density
\begin{equation}
\tilde{\rho}=\frac{k^{2}d^{2}}{(2k-1)\tilde{M}^{2}}\rho
\end{equation}
as a function of $\tilde{r}=r/d$ and $\tilde{z}=z/d$. We can observe two kinds of behavior, both of them present in all the particular solutions discussed in this paper, that can be obtained by choosing properly the constant values $k$, $\tilde{M}$. The first behavior is the one presented in the plots for $m=2,3$, where the energy density reaches its maximum at the center of the halo ($\tilde{z}\approx\tilde{r}\approx0$), and then goes to zero at infinity. It is clear, in this case, that the energy is more concentrated in the region near the disk. The second behavior is shown in the plot associated to the $m=1$ model. Here we see an unusual distribution of energy (if we think, only, in the visible matter of the halo), where the maximum of density is located at $\tilde{z}\approx0$ and $\tilde{r}\approx1$, i.e. near the edge of disk, and goes to zero at infinity.

We can also get analytic expressions for the energy density of the disk for the particular models considered in this section by using (\ref{uobla}) in (\ref{eq20s}), so that
\begin{subequations}\label{s}
\begin{eqnarray}
\tilde{\sigma}_{1}&=&\frac{12(1-\tilde{r}^2)^{1/2}}{\left[1+\frac{3\pi}{8}\tilde{M}A_{1}(\tilde{r})\right]^{\frac{1+k}{k}}},\label{s1}\\ 
\tilde{\sigma}_{2}&=&\frac{20(1-\tilde{r}^{2})^{3/2}}{\left[1+\frac{15\pi}{128}\tilde{M}A_{2}(\tilde{r})\right]^{\frac{1+k}{k}}},\label{s2}\\ 
\tilde{\sigma}_{3}&=&\frac{28(1-\tilde{r}^{2})^{5/2}}{\left[1+\frac{35\pi}{512}\tilde{M}A_{3}(\tilde{r})\right]^{\frac{1+k}{k}}},\label{s3}
\end{eqnarray}
\end{subequations}\\
where $\tilde{\sigma}_{m}=(kd/\tilde{M})\sigma_{m}$, $\tilde{r}=r/d$, $\tilde{M}=M/d$, and
\begin{subequations}
\begin{eqnarray}
A_{1}(\tilde{r})&=&2-\tilde{r}^{2},\\ 
A_{2}(\tilde{r})&=&8-8\tilde{r}^{2}+3\tilde{r}^{4},\\ 
A_{3}(\tilde{r})&=&16-24\tilde{r}^{2}+18\tilde{r}^{4}-5\tilde{r}^{6}.
\end{eqnarray}
\end{subequations}

\begin{figure*}
$$\begin{array}{cc}
m = 1, 2, 3 & m = 1 \\
\epsfig{width=3in,file=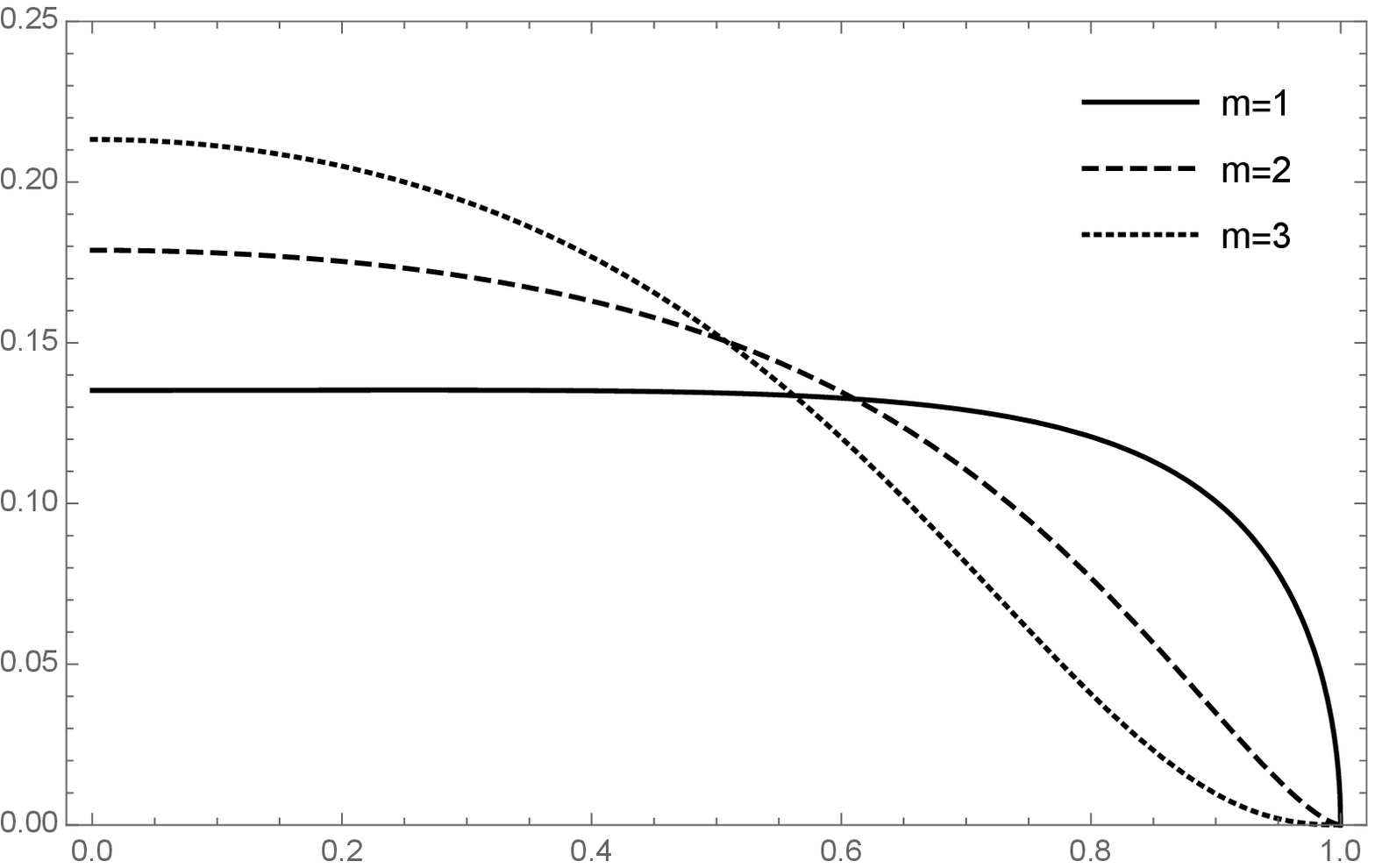} &
\epsfig{width=3in,file=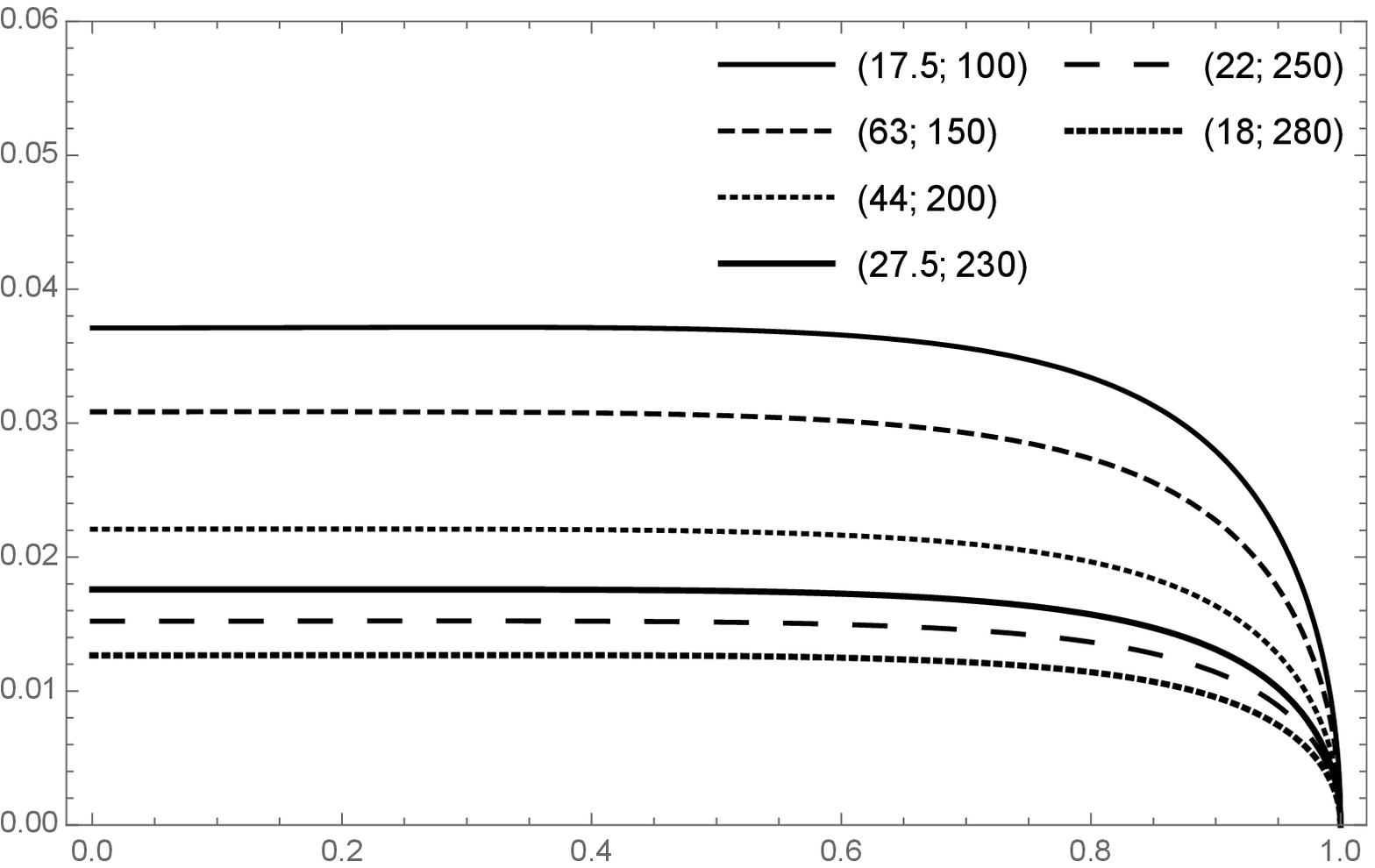} \\ \\
m = 2 & m = 3 \\
\epsfig{width=3in,file=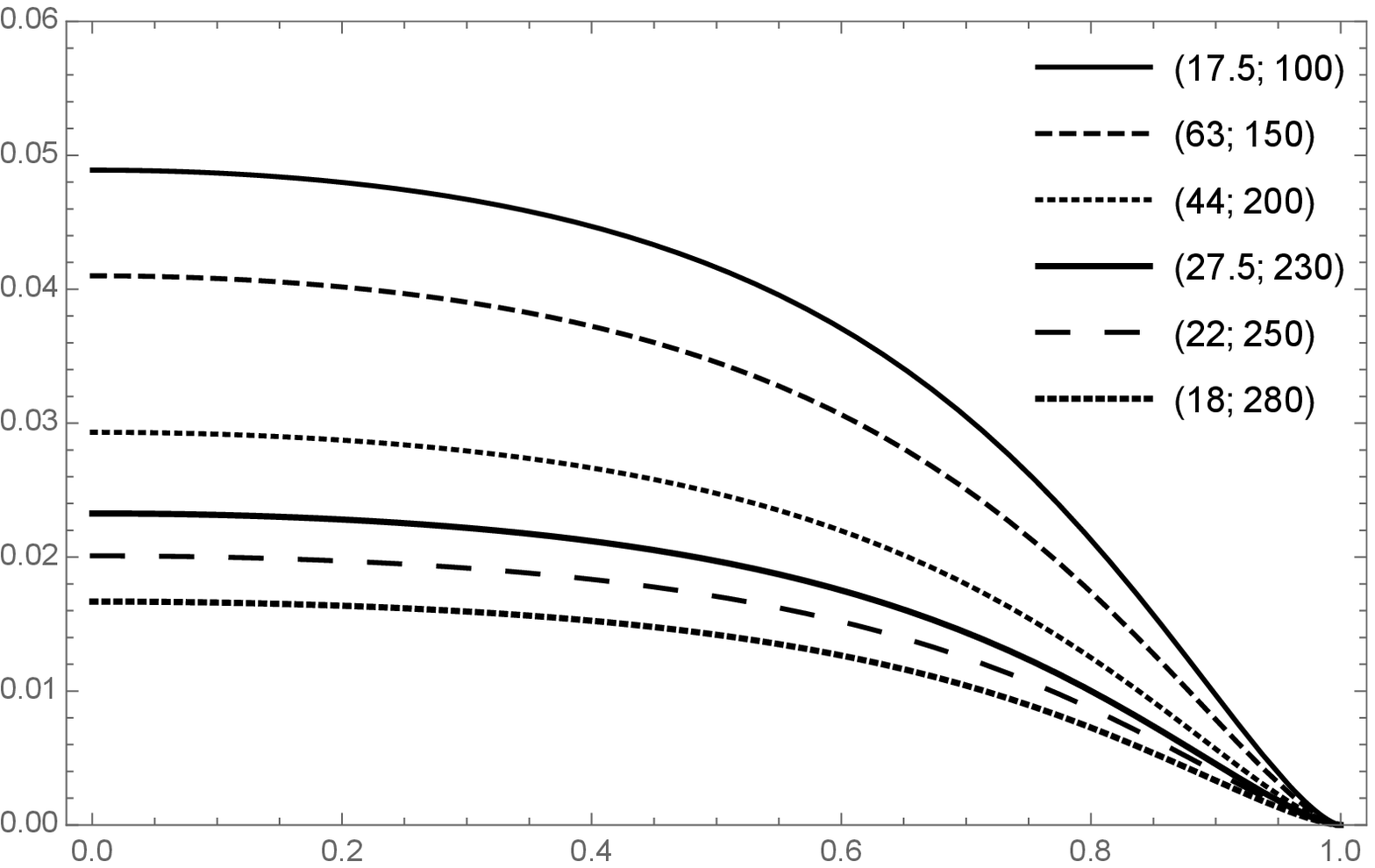} &
\epsfig{width=3in,file=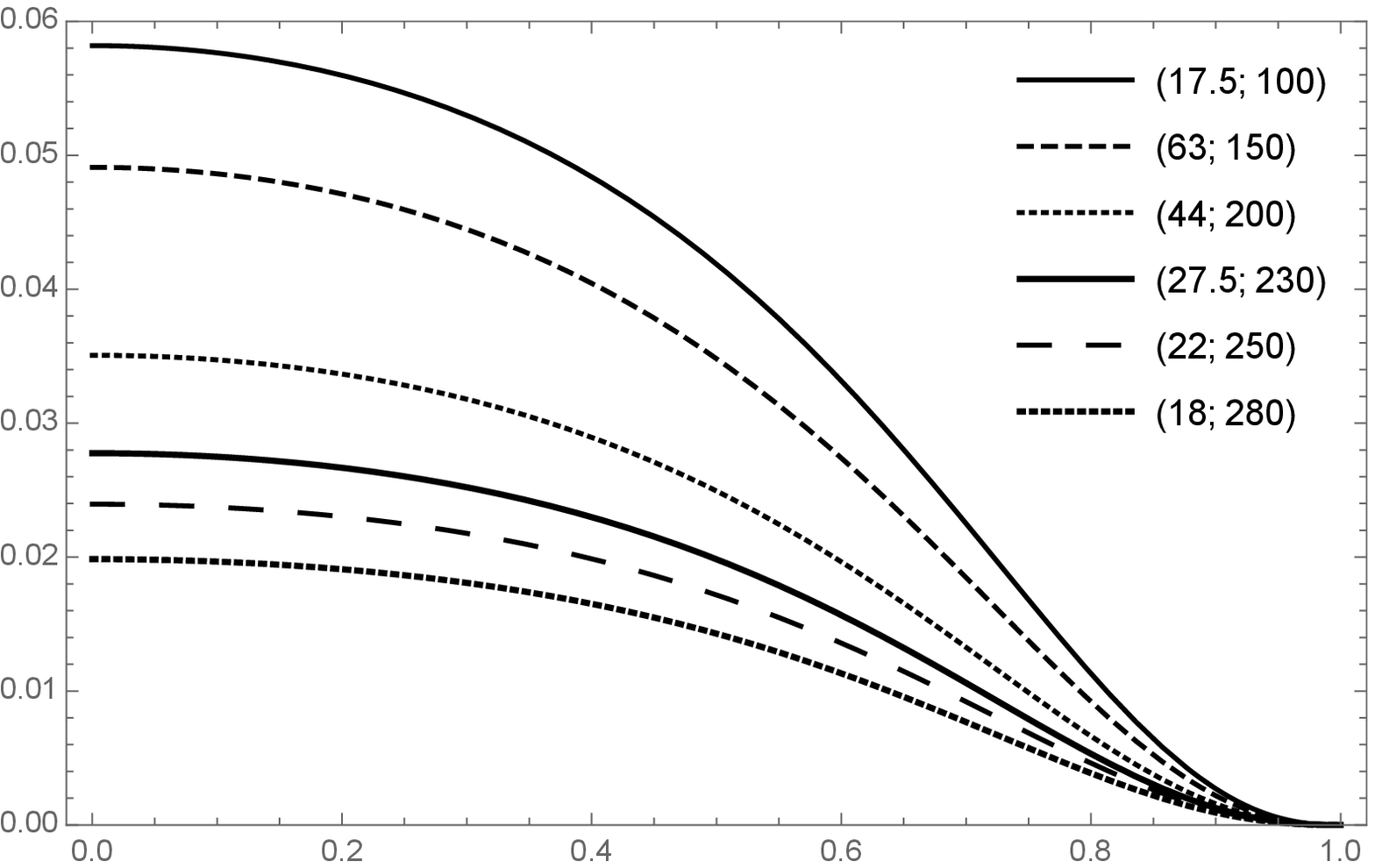} \\
\end{array}$$
\caption{\label{dendiskal}Surface energy density of the disk $\tilde{\sigma}_{n}$ as a function of $\tilde{r}$ for the first 3 models of the family of solutions. The first plot (top-left) correspond to the three models with the same $k=20$ and $\tilde{M}=30$. In the other plots $m$ is kept constant while $k$ and $\tilde{M}$ are changed; each curve in these plots is labeled by a pair of numbers ($k, \tilde{M}$)}
\end{figure*}

In FIG. \ref{dendiskal}, we plot the energy density $\tilde{\sigma}$ of the disk as a function of the normalized radius for the models $m=1,2,3$. In the first plot we compare the energy density for the constants values $k=20$ and $\tilde{M}=30$, and for different values of $m$. In this plot we see that $\tilde{\sigma}$ has a maximum at the center of the system $(\tilde{r}=0)$, and goes to zero at the edge ($\tilde{r}=1$). The spatial distribution of the energy depends on the particular value of $m$: for $m=1$, we see that the energy density remains approximately constant, and starts to fall from $\tilde{r}\approx 0.6$; but if we increase the value of $m$, then the energy density will be more concentrated at the center, and its maximum value will be higher. The remaining plots show how, by changing $k$ and $\tilde{M}$, we can get different kinds of behavior in each particular model. It is clear that we can manipulate the maximum value of the energy density and its rate of decline by varying the constants, but in all cases the behavior is very similar and the energy is concentrated in the center of the disk.

Now, with (\ref{eq8}) we can write the gravitational mass (\ref{Mh}) equivalently as
\begin{equation}
M=\frac{2}{k}\oint\frac{\nabla U}{1-U}\cdot d\vec{s},
\label{Mheq}
\end{equation}
where $s$ is the area of a closed surface which contains a mass $M$. We choose $s$ to be the surface of an ellipsoid determined by $\xi$ constant. Accordingly, we can define the mass of the disk throw the integral
\begin{equation}
M_{D}=\frac{2a}{k}\int_{0}^{2\pi}\int_{-1}^{1}\left.\frac{(1+\xi^2)U_{,\xi}}{1-U}\right|_{\xi=0}d\eta d\varphi,
\label{Mdobla}
\end{equation}
and the total mass of the system by
\begin{equation}
M_{T}=\frac{2a}{k}\int_{0}^{2\pi}\int_{-1}^{1}\left.\frac{(1+\xi^2)U_{,\xi}}{1-U}\right|_{\xi=\infty}d\eta d\varphi,
\label{Mtobla}
\end{equation}
so that, the mass of the halo is $M_{H}=M_{T}-M_{D}$.\\

Despite the fact that in this model the halo has infinite extension, it is possible to compute the total mass of the system through the integral given on (\ref{Mtobla}), and show that $M_{T}$ converges. To do it, we first take advantage of the fact that at infinity, $U$ tends to zero. So, by using (\ref{uobla}), we find that
\begin{equation}
(1+\xi^{2})U_{,\xi} = C_{0}-\sum_{n=1}^{\infty}C_{2n}(1+\xi^{2}) q^\prime_{2n} (\xi) P_{2n}(\eta).
\label{eq25-1}
\end{equation}
Now, with the recurrence relation,
\begin{equation}
(1+\xi^{2}) q^\prime_{n} (\xi) = n\xi q_{n}(\xi)-nq_{n-1}(\xi),
\label{eq26-1}
\end{equation}
we can transform (\ref{eq25-1}) on
\begin{equation}
(1+\xi^{2})U_{,\xi}=C_{0}-\sum_{n=1}^{\infty}C_{2n}(2n\xi q_{2n}-2nq_{2n-1})P_{2n}.
\label{eq27-1}
\end{equation}
As $q_{2n-1}(\xi)$ becomes zero at infinity, and $\xi q_{2n}(\xi)$ goes to $0$ if $\xi\to\infty$ for all $n\geq1$, then $(1+\xi^{2})U_{,\xi}|_{_{\xi=\infty}}=C_{0}$, and the total mass of the disk-halo system becomes
\begin{equation}
M_{T_{m}}=\frac{8\pi dC_{0}}{k},
\label{Mtall}
\end{equation}
which depends on the radius of the disk, the first constant $C_{0}$ and the particular kind of fluid in the halo (determined by the value of $k$). With this, we proof that the total mass is finite, even though the halo has infinite extension. This proof is very general, so $M_{T}$ is the same for all particular models. Now, for the first model ($m=1$) the integral (\ref{Mdobla}) yield to
\begin{equation}
M_{D_{1}} = \frac{64d}{k} \left[ 1 - \sqrt{1 + \frac{8}{3 \pi C_{0}}} \acot\sqrt{\frac{3 \pi C_{0}}{3 \pi C_{0} + 8}} \right],
\label{eq36}
\end{equation}
where $C_{0}=M/d$; so the mass of the relativistic disk is related to the newtonian mass.

The analytic expression for the circular velocity of a particle on the plane $z=0$ has two parts: one part correspond to the particle moving on the disk ($0\leq \tilde{r}\leq 1$), for which we need to do $\xi=0$ on (\ref{eq33}); and the other part correspond to the velocity of a particle moving outside the disk ($\tilde{r}\geq 1$), for which we have to do $\eta=0$ on the expression for $v_{c}^{2}$. Both parts must coincide exactly at $\tilde{r}=1$. Now, by using the expressions for $U_{m}$ and $U_{m,r}$ in (\ref{eq33}), and doing the appropriate evaluation in each region we get the solutions for the circular velocity for the models we are considering, which are given by
\begin{widetext}
\begin{align}
v_{c_{1}}^{2} &= \left\{
\begin{array}{ll}
\dfrac{6 \tilde{M} \pi \tilde{r}^{2}}{k [8 - 3 \tilde{M} \pi (\tilde{r}^{2} - 2) ] - 6 \tilde{M} \pi \tilde{r}^{2}},  & 0 \leq \tilde{r} \leq 1, \\
             \\
\dfrac{6 \tilde{M} ( \tilde{r}^{2} \acot J - J )}{6 \tilde{M} J + k (4 + 3 \tilde{M} J) - 3 \tilde{M} [2 \tilde{r}^{2} + k (\tilde{r}^{2} - 2) ] \acot J}, & \tilde{r} \geq 1,
\end{array} \right. \\
\nonumber \\
v_{c_{2}}^{2} &= \left\{
\begin{array}{ll}
\dfrac{- 60 \tilde{M} \pi \tilde{r}^{2} L_{1}}{60 \tilde{M} \pi \tilde{r}^{2}L_{1} + k (128 + 15 \tilde{M} \pi L_{2})}, & 0 \leq \tilde{r} \leq 1, \\
             \\
\dfrac{60 \tilde{M} [\tilde{r}^{2}L_{1} \acot J - J L_{3} ]}{60\tilde{M} J L_{3} + k [45\tilde{M}(\tilde{r}^{2} - 2) J - 64 ] - 15 \tilde{M} (4 \tilde{r}^{2} L_{1} + k L_{2}) \acot J}, & \tilde{r}\geq 1,
\end{array} \right. \\
\nonumber \\
v_{c_{3}}^{2} &= \left\{
\begin{array}{ll}
\dfrac{-210 \tilde{M} \pi \tilde{r}^{2} N_{1}}{210 \tilde{M} \pi \tilde{r}^{2}N_{1} + k (35\tilde{M}\pi N_{2} - 512)}, &  0\leq \tilde{r}\leq 1, \\
             \\
\dfrac{210 \tilde{M} [3 \tilde{r}^{2} N_{1} \acot J - J N_{3} ]}{210 \tilde{M} J N_{3} + k (768 + 35 \tilde{M} J N_{4}) - 105 \tilde{M} (6 \tilde{r}^{2}N_{1} + k N_{2}) \acot J}, &  \tilde{r}\geq 1,
\end{array} \right.
\end{align}
where $J = \sqrt{\tilde{r}^{2}-1}$, $L_{1}=-4+3\tilde{r}^{2}$, $L_{2}=8-8\tilde{r}^{2} + 3 \tilde{r}^{4}$, $L_{3} = - 2 + 3 \tilde{r}^{2}$, $N_{1}=8-12\tilde{r}^{2} + 5\tilde{r}^{4}$, $N_{2}= - 16 + 24 \tilde{r}^{2}-18\tilde{r}^{4}+5\tilde{r}^{6}$, $N_{3}=8-26\tilde{r}^{2}+15\tilde{r}^{4}$ and $N_{4}=44-44\tilde{r}^{2}+15\tilde{r}^{4}$.
\end{widetext}

\begin{figure*}
$$\begin{array}{cc}
m = 1, 2, 3 & m = 1 \\
\epsfig{width=3in,file=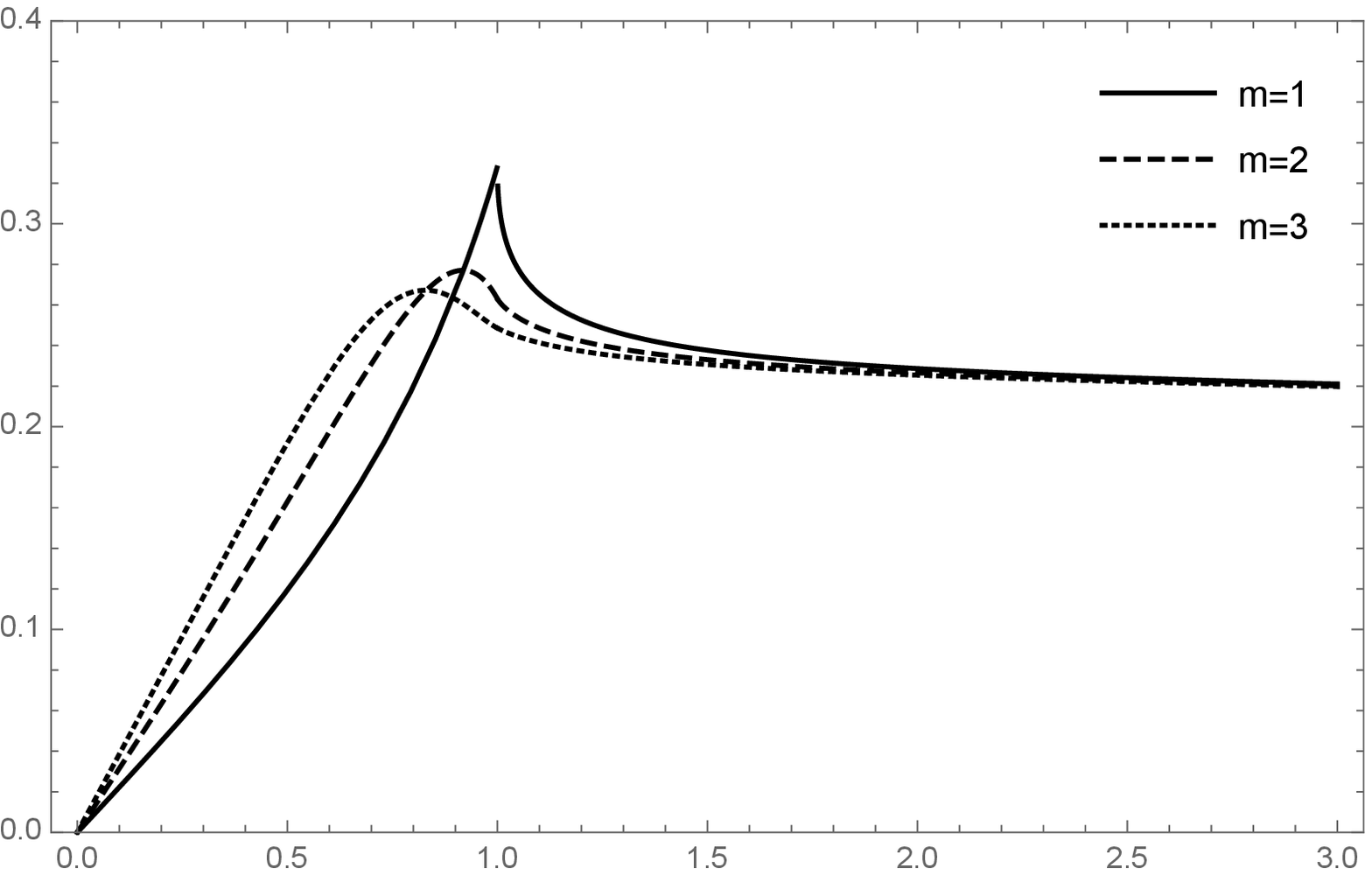} &
\epsfig{width=3in,file=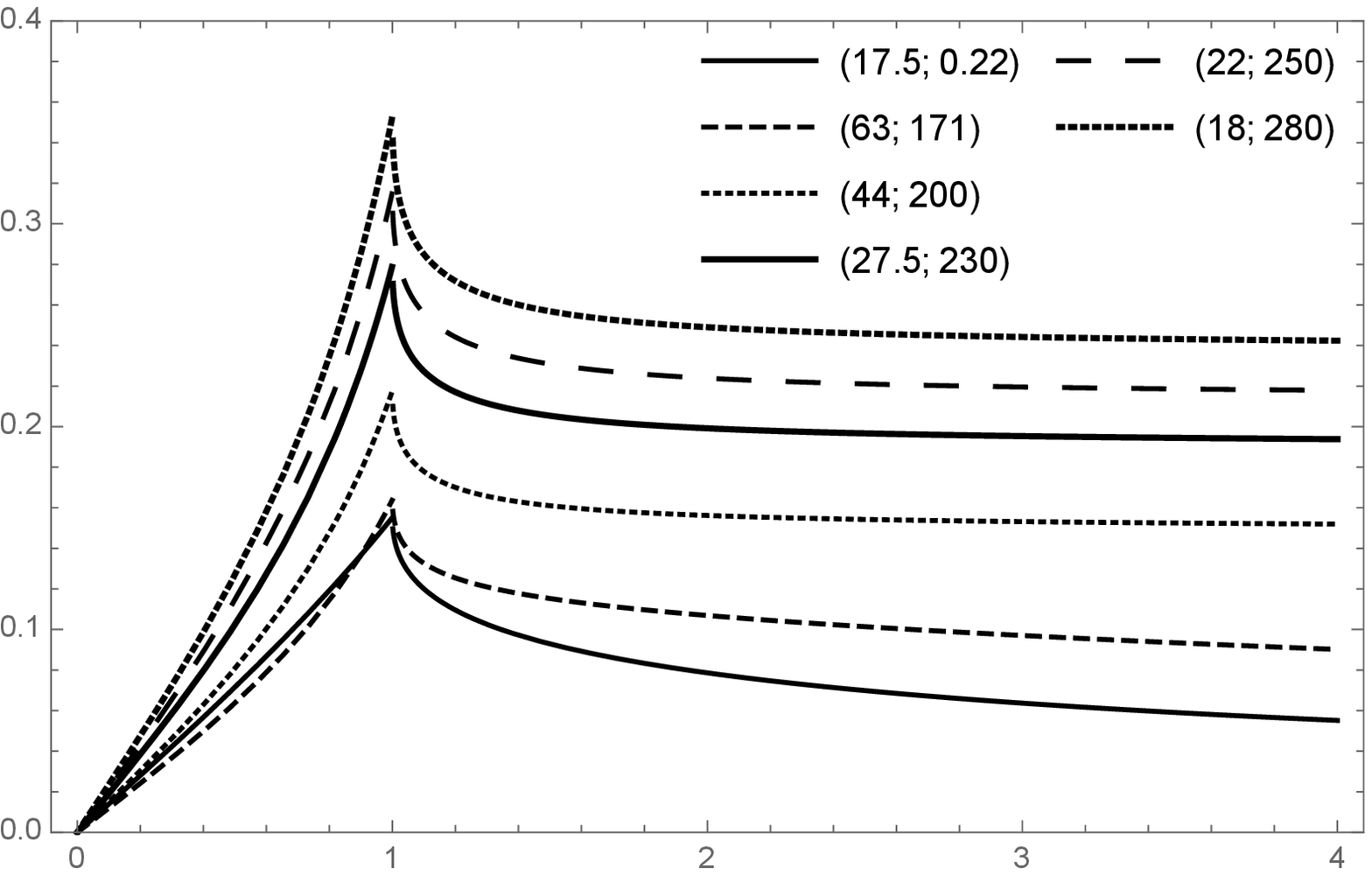} \\ \\
m = 2 & m = 3 \\
\epsfig{width=3in,file=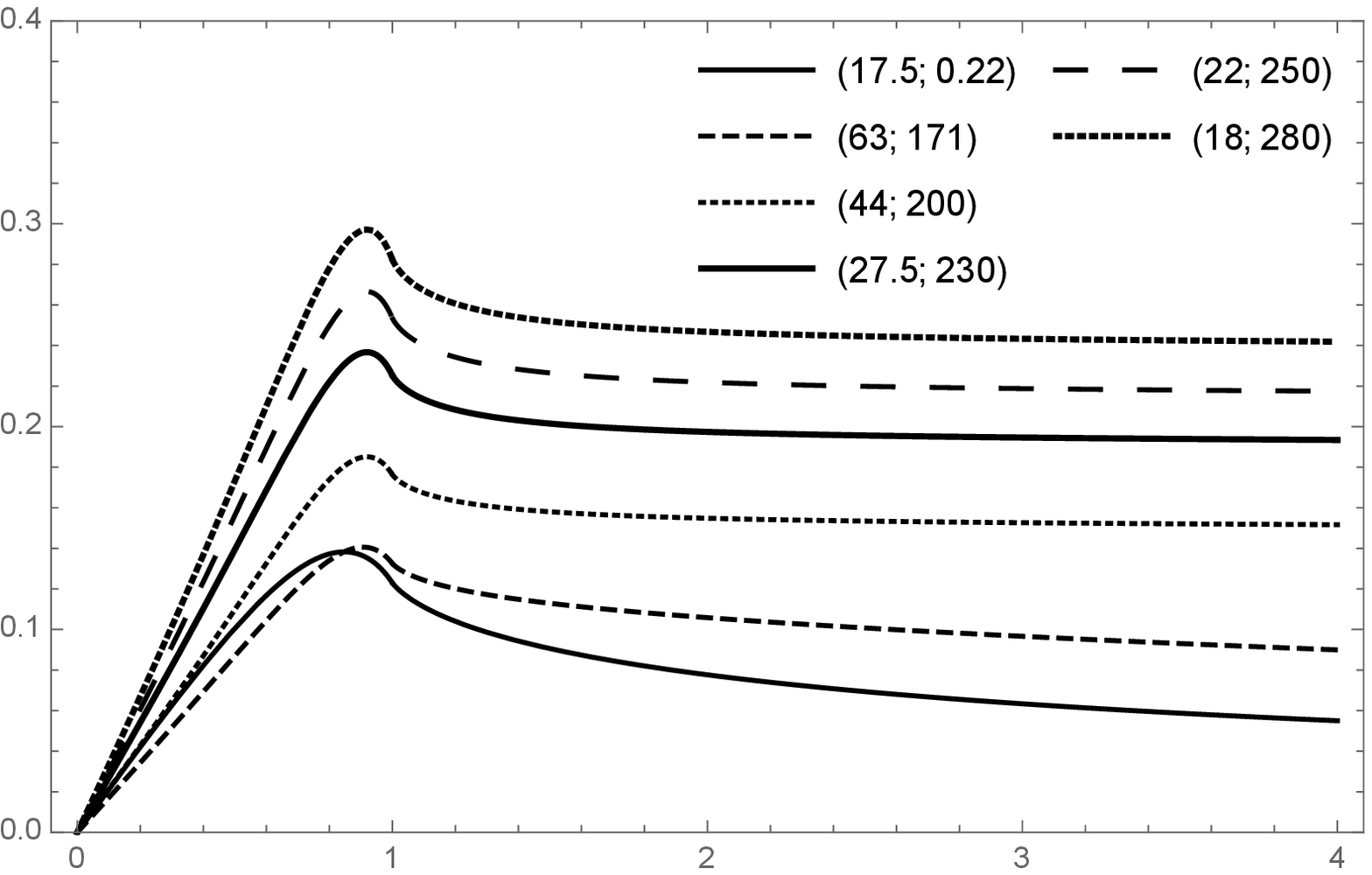}  &
\epsfig{width=3in,file=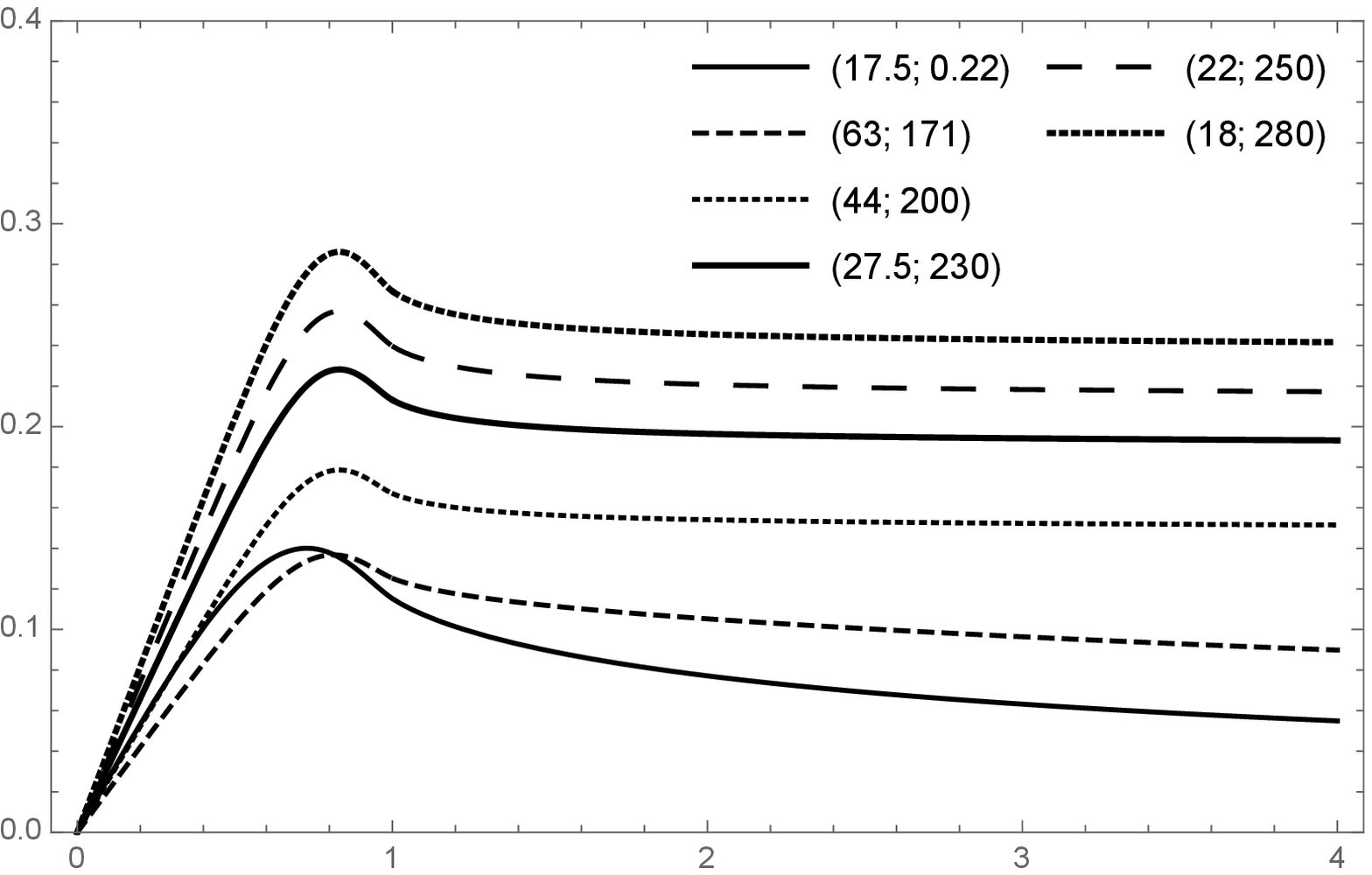} \\
\end{array}$$
\caption{\label{velcirckal}Circular velocity $v_{c_{m}}$ as a function of $\tilde{r}$ for the first 3 models of the family of solutions. The first plot (top-left) correspond to the three models with the same $k=20$ and $\tilde{M}=30$. In the other plots $m$ is kept constant while $k$ and $\tilde{M}$ are changed; each curve in these plots is labeled by a pair of numbers ($k, \tilde{M}$)}
\end{figure*}

In FIG. \ref{velcirckal}, we show the profiles of the circular velocity as a function of $\tilde{r}$ for some constants values $k, \tilde{M}$. All the constants values, in this paper, are chosen in such a way that the physical quantities are well behaved ($\tilde{\sigma}_{m}\geq 0$ and $0\leq v_{c}\leq 1$). In the first plot, in FIG. \ref{velcirckal}, we compare the rotational curve in the models $m=1,2,3$ for $k$ and $\tilde{M}$ fixed. Here we can see that if $m$ is higher, then the peak would be less acute, the union between the two curves mentioned in the last paragraph would be smoother and the maximum value of $v_{c}$ would be attained closer to the center of the disk.\\

In all particular solutions, the behavior of $v_{c}$ is the same: the velocity starts from zero and reaches a maximum that depends on the particular constants values, then the velocity falls to a value that remains about constant; this is very important because many real rotational curves from spiral galaxies \cite{26} has the same behavior (qualitatively speaking). An specific shape for the rotational curve can be obtained by choosing properly the values for $m, k$ and $\tilde{M}$, e.g, in the plots for $m=1,2,3$, the curves associated to the values $k=17.5$ and $\tilde{M}=0.22$ show that we can also get circular velocities that fall after reaching the maximum.

\section{\label{sec:concluding}Concluding Remarks}

We obtained two sets of new exact solutions to the Einstein equations for an axially symmetric static spacetime from the conformastatic metric. These solutions describe the physical properties of an infinite disk of dust, i.e. without stresses, which is surrounded by a fluid or halo with an equation of state $p=[3(2k-1)]^{-1}\rho$. This equation implies different kind of fluids: radiation if $k=1$, dust if $k$ tends to infinity or another kind of fluid according to the value of $k$. The disk-like solutions obtained here are physically well behaved because they satisfy the energy conditions required for any source of gravitational field, and because they are asymptotically flat. The asymptotic behavior was ensured by writing an appropriate relation between the metric function $\psi$ and the auxiliary function $U$. 

According to the solutions, we got for the halo a fluid whose main stresses are different from zero ($p_{1}=p_{2}=-p_{3}\neq 0$) and they are in the same direction of the main axes. On the contrary, the solutions obtained for the disk describe a fluid with null stress. In the Kuzmin-Toomre disks with halo we obtained that the most of the energy density was located at the center of the system; so it is possible to see a well-defined central region with a maximum at ($\tilde{r}=0$ and $\tilde{z}=0$); nevertheless, as far as $n$ increases, the amount of constants increases too, so we have more free parameters to adjust the energy density profile to a particular behavior. For instance, we presented in FIG. \ref{densidiskuz} one case in which the maximum energy density is not at the center of the system. We found that the Morgan-Morgan relativistic thin disks models presented in Section \ref{sec:morgan} does not have the same freedom as the Kuzmin-Toomre ones. In these models we only have two parameters to adjust: $k$ and $\tilde{M}$, so by varing these parameters we only change the maximum of $\sigma$, $\rho$ and $v_{c}$, and their corresponding rates of decrease.

We computed the total mass for the $n=0$ model of the Kuzmin-Toomre relativistic disks with halo, turning out that $M_{T_{0}}=\frac{8\pi A_{0}}{k}$, and we proved that the mass of the system converges for all $n$. On the other hand, in the Morgan-Morgan relativistic disks solutions we computed the total gravitational mass for all the family of solutions which is $M_{T_{m}}=\frac{8\pi dC_{0}}{k}$. Finally, we presented the rotational curves for some particular models in both families of solutions. We saw that, for some values of the constants, it is possible to get a behavior in which the square of the circular velocity increases from $\tilde{r}=0$ until a radius where $v_{c}$ stays approximately constant. Nevertheless, we got some rotational curves with two peaks in the model $n=3$ of the Kuzmin-Toomre relativistic solutions. This is a very interesting behavior because the observed rotational curves in spiral galaxies exhibit these kind of features \cite{26}.\\

\section*{Acknowledgments}

GAG was supported in part by VIE-UIS, under grants number 1347 and 1838, and COLCIENCIAS, Colombia, under grant number 8840. OMP wants to thank the financial support from the program {\em J\'ovenes Investigadores} of COLCIENCIAS, Colombia.

\end{document}